\documentclass{ieeeaccess}
\usepackage{amsmath,amssymb,amsfonts}
\usepackage{algorithmic}
\usepackage{graphicx}
\usepackage{textcomp}
\usepackage{multirow}

\usepackage[backend=bibtex,natbib=true,sorting=none]{biblatex}
\usepackage{xspace} 
\usepackage{upgreek}

\def\MagUp {\mbox{\em Mag\kern -0.05em Up}\xspace}

{

 \def\PDelta      {\ensuremath{\Delta}\xspace}                 
 \def\PXi         {\ensuremath{\Xi}\xspace}                 
 \def\PLambda     {\ensuremath{\Lambda}\xspace}                 
 \def\PSigma      {\ensuremath{\Sigma}\xspace}                 
 \def\POmega      {\ensuremath{\Omega}\xspace}                 
 \def\PUpsilon    {\ensuremath{\Upsilon}\xspace}

 \def\PB      {\ensuremath{\mathrm{B}}\xspace}                 
                  
 \def\PD      {\ensuremath{\mathrm{D}}\xspace}

 \def\PK      {\ensuremath{\mathrm{K}}\xspace}

 \def\Pi      {\ensuremath{\mathrm{i}}\xspace}

 \def\Ps      {\ensuremath{\mathrm{s}}\xspace}

 \def\thebaroffset{0.0em}
}
{

 \mathchardef\PDelta="7101
 \mathchardef\PXi="7104
 \mathchardef\PLambda="7103
 \mathchardef\PSigma="7106
 \mathchardef\POmega="710A
 \mathchardef\PUpsilon="7107
                  
 \def\PB      {\ensuremath{B}\xspace}                 
                  
 \def\PD      {\ensuremath{D}\xspace}

 \def\PK      {\ensuremath{K}\xspace}

 \def\Pi      {\ensuremath{i}\xspace}

 \def\Ps      {\ensuremath{s}\xspace}

 \def\thebaroffset{0.18em}
}
\newcommand{\offsetoverline}[2][\thebaroffset]{\kern #1\overline{\kern -#1 #2}}%

\makeatletter
\ifcase \@ptsize \relax
  \newcommand{\miniscule}{\@setfontsize\miniscule{4}{5}}
\or
  \newcommand{\miniscule}{\@setfontsize\miniscule{5}{6}}
\or
  \newcommand{\miniscule}{\@setfontsize\miniscule{5}{6}}
\fi
\makeatother

\DeclareRobustCommand{\optbar}[1]{\shortstack{{\miniscule (\rule[.5ex]{1.25em}{.18mm})}
  \\ [-.7ex] $#1$}}

\def\squark    {{\ensuremath{\Ps}}\xspace}

\def\KorKbar {\kern \thebaroffset\optbar{\kern -\thebaroffset \PK}{}\xspace}

\def\D       {{\ensuremath{\PD}}\xspace}

\def\DorDbar {\kern \thebaroffset\optbar{\kern -\thebaroffset \PD}\xspace}

\def\Dp      {{\ensuremath{\D^+}}\xspace}
\def\Dm      {{\ensuremath{\D^-}}\xspace}

\def\DpDm    {\ensuremath{\Dp {\kern -0.16em \Dm}}\xspace}

\def\B       {{\ensuremath{\PB}}\xspace}

\def\BorBbar {\kern \thebaroffset\optbar{\kern -\thebaroffset \PB}\xspace}

\def\Bd      {{\ensuremath{\B^0}}\xspace}

\def\BdorBdbar {\kern \thebaroffset\optbar{\kern -\thebaroffset \Bd}\xspace}

\def\Bs      {{\ensuremath{\B^0_\squark}}\xspace}

\def\BsorBsbar {\kern \thebaroffset\optbar{\kern -\thebaroffset \Bs}\xspace}

\def\Y#1S{\ensuremath{\PUpsilon{(#1S)}}\xspace}

\def\LorLbar     {\kern \thebaroffset\optbar{\kern -\thebaroffset \PLambda}\xspace}

\def\to                 {\ensuremath{\rightarrow}\xspace}

\def\AT#1     {\ensuremath{A_{\mathrm{T}}^{#1}}\xspace}           

\def\C#1      {\ensuremath{\mathcal{C}_{#1}}\xspace}                       
\def\Cp#1     {\ensuremath{\mathcal{C}_{#1}^{'}}\xspace}                    
\def\Ceff#1   {\ensuremath{\mathcal{C}_{#1}^{\mathrm{(eff)}}}\xspace}        
\def\Cpeff#1  {\ensuremath{\mathcal{C}_{#1}^{'\mathrm{(eff)}}}\xspace}       
\def\Ope#1    {\ensuremath{\mathcal{O}_{#1}}\xspace}                       
\def\Opep#1   {\ensuremath{\mathcal{O}_{#1}^{'}}\xspace}                    

\newcommand{\aunit}[1]{\ensuremath{\text{\,#1}}}       

\newcommand{\tev}{\aunit{Te\kern -0.1em V}\xspace}
\newcommand{\gev}{\aunit{Ge\kern -0.1em V}\xspace}
\newcommand{\mev}{\aunit{Me\kern -0.1em V}\xspace}
\newcommand{\kev}{\aunit{ke\kern -0.1em V}\xspace}
\newcommand{\ev}{\aunit{e\kern -0.1em V}\xspace}
 
\newcommand{\mevc}{\ensuremath{\aunit{Me\kern -0.1em V\!/}c}\xspace}
\newcommand{\gevc}{\ensuremath{\aunit{Ge\kern -0.1em V\!/}c}\xspace}
\newcommand{\mevcc}{\ensuremath{\aunit{Me\kern -0.1em V\!/}c^2}\xspace}
\newcommand{\gevcc}{\ensuremath{\aunit{Ge\kern -0.1em V\!/}c^2}\xspace}

\def\gsim{{~\raise.15em\hbox{$>$}\kern-.85em
          \lower.35em\hbox{$\sim$}~}\xspace}
\def\lsim{{~\raise.15em\hbox{$<$}\kern-.85em
          \lower.35em\hbox{$\sim$}~}\xspace}

\def\pt         {\ensuremath{p_{\mathrm{T}}}\xspace}

\def\tell1  {TELL1\xspace}
\def\ukl1   {UKL1\xspace}

\def\BibTeX{{\rm B\kern-.05em{\sc i\kern-.025em b}\kern-.08em
    T\kern-.1667em\lower.7ex\hbox{E}\kern-.125emX}}
\begin{document}
\history{Date of publication xxxx 00, 0000, date of current version xxxx 00, 0000.}
\doi{xx.xxxx/ACCESS.2024.DOI}

\title{Looking Forward: A High-Throughput Track Following Algorithm for Parallel Architectures}
\author{\uppercase{Aurelien Bailly-Reyre}\authorrefmark{1}, \uppercase{Lingzhu Bian}\authorrefmark{2},
\uppercase{Pierre Billoir}\authorrefmark{1}, \uppercase{Daniel Hugo C\'ampora P\'erez}\authorrefmark{3}, 
\uppercase{Vladimir Vava Gligorov}\authorrefmark{1}, \uppercase{Flavio Pisani}\authorrefmark{4}, 
\uppercase{Renato Quagliani}\authorrefmark{1,4,5}, \uppercase{Alessandro Scarabotto}\authorrefmark{1,6} and \uppercase{Dorothea vom Bruch}\authorrefmark{7}}
\address[1]{LPNHE, Sorbonne Universit{\'e}, Paris Diderot Sorbonne Paris Cit{\'e}, CNRS/IN2P3}
\address[2]{Wuhan University}
\address[3]{Universiteit Maastricht}
\address[4]{European Organization for Nuclear Research (CERN)}
\address[5]{Institute of Physics, Ecole Polytechnique F{\'e}d{\'e}rale de Lausanne (EPFL)}
\address[6]{Technische Universit{\"a}t (TU) Dortmund}
\address[7]{Aix Marseille Univ, CNRS/IN2P3, CPPM}

\tfootnote{RQ is now based at CERN [4]. AS is now based in TU Dortmund [6].}

\markboth
{Author \headeretal: Preparation of Papers for IEEE TRANSACTIONS and JOURNALS}
{Author \headeretal: Preparation of Papers for IEEE TRANSACTIONS and JOURNALS}

\corresp{Corresponding authors: D.H. C\'ampora P\'erez (e-mail: dcampora@cern.ch), V. Gligorov (email: vladimir.gligorov@cern.ch), R. Quagliani (e-mail: renato.quagliani@cern.ch) and A. Scarabotto (email: alessandro.scarabotto@cern.ch).}

\begin{abstract}
Real-time data processing is a central aspect of particle physics experiments with high requirements on computing resources. The LHCb experiment must cope with the 30 million proton-proton bunches collision per second rate of the Large Hadron Collider (LHC), producing $10^9$ particles/s. The large input data rate of 32 Tb/s needs to be processed in real time by the LHCb trigger system, which includes both reconstruction and selection algorithms to reduce the number of saved events. The trigger system is implemented in two stages and deployed in a custom data centre.

We present Looking Forward, a high-throughput track following algorithm designed for the first stage of the LHCb trigger and optimised for GPUs. The algorithm focuses on the reconstruction of particles traversing the whole LHCb detector and is developed to obtain the best physics performance while respecting the throughput limitations of the trigger. The physics and computing performances are discussed and validated with simulated samples. 
\end{abstract}


\titlepgskip=-15pt

\maketitle

\section{Introduction}
\label{sec:introduction}
The real-time or near-real-time reconstruction of charged particle trajectories (tracking) has been a central element of 
detectors at hadron colliders since the UA1 experiment at CERN (1981-1990)~\cite{Dorenbosch:1985cx}. The rate and complexity of particle collisions have 
increased over the past decades and so have the computational demands placed on real-time tracking algorithms. This 
motivates continued research into high-throughput tracking algorithms which can efficiently exploit modern parallel 
computing architectures. Tracking algorithms consist in associating together hits left by charged particles in the detector to form tracks. These hits are then fitted to a track model in order to extract kinematic and geometric properties.
Reconstruction algorithms are generally one of the most time-consuming components of data processing
pipelines, which reconstruct physics 
quantities of interest with the highest possible precision and fidelity compared to the already recorded data to permanent storage. The need to reduce the computational cost 
and energy consumption of offline processing therefore motivates research into efficient tracking algorithms. In the 
past, experiments often deployed special real-time algorithms tuned to specific real-time processing architectures~\cite{LHCb:2018zdd,ATLAS:2020esi,CMS:2016ngn} 
which were decoupled from the offline algorithms and architectures. Increasingly, however, both real-time and
offline processing is carried out in data centres populated by heterogeneous computing architectures. It is therefore 
important in both cases to carefully benchmark, in as general a manner as possible, the tradeoffs between physics performance 
and computational scalability for different architectures. 

In this paper we present ``Looking Forward'', a high-throughput algorithm for reconstructing charged tracks in the LHCb~\cite{LHCb:2023hlw} 
detector during Run~3 of the Large Hadron Collider (LHC) which is taking place from 2022 to 2025 inclusive, achieving an instantaneous luminosity of ${\mathcal{L} = 2 \times 10^{33} cm^{-2} s^{-1}}$, five times larger compared to Run~2. The computing resources of LHCb require that the input data rate of 32~Tb/s is reduced to 10~GB/s using selections which are primarily 
based on the properties of reconstructed tracks. In order to achieve this goal, LHCb uses a two-stage real-time processing 
pipeline deployed in a custom data centre~\cite{Colombo:2019bel}. The first stage is executed entirely on GPU processors, after which the 
data is buffered and the best fidelity detector alignment and calibration are deployed. Subsequently, a second stage is 
executed entirely on CPU processors. This division of labour between CPU and GPU processors has been optimised~\cite{LHCb:2023hlw} for 
Run~3 conditions, but may evolve in the future. The Looking Forward algorithm is therefore designed for deployment on both 
GPU and CPU architectures. Furthermore, it is an excellent case study of the more 
general kind of algorithmic optimisation motivated earlier. The scope of the algorithm is to reconstruct tracks traversing the whole LHCb detector which are fundamental when triggering on interesting events at LHCb.

\section{Context}
\label{sec:context}
\begin{figure}[t]
    \centering
    \includegraphics[width=0.45\textwidth]{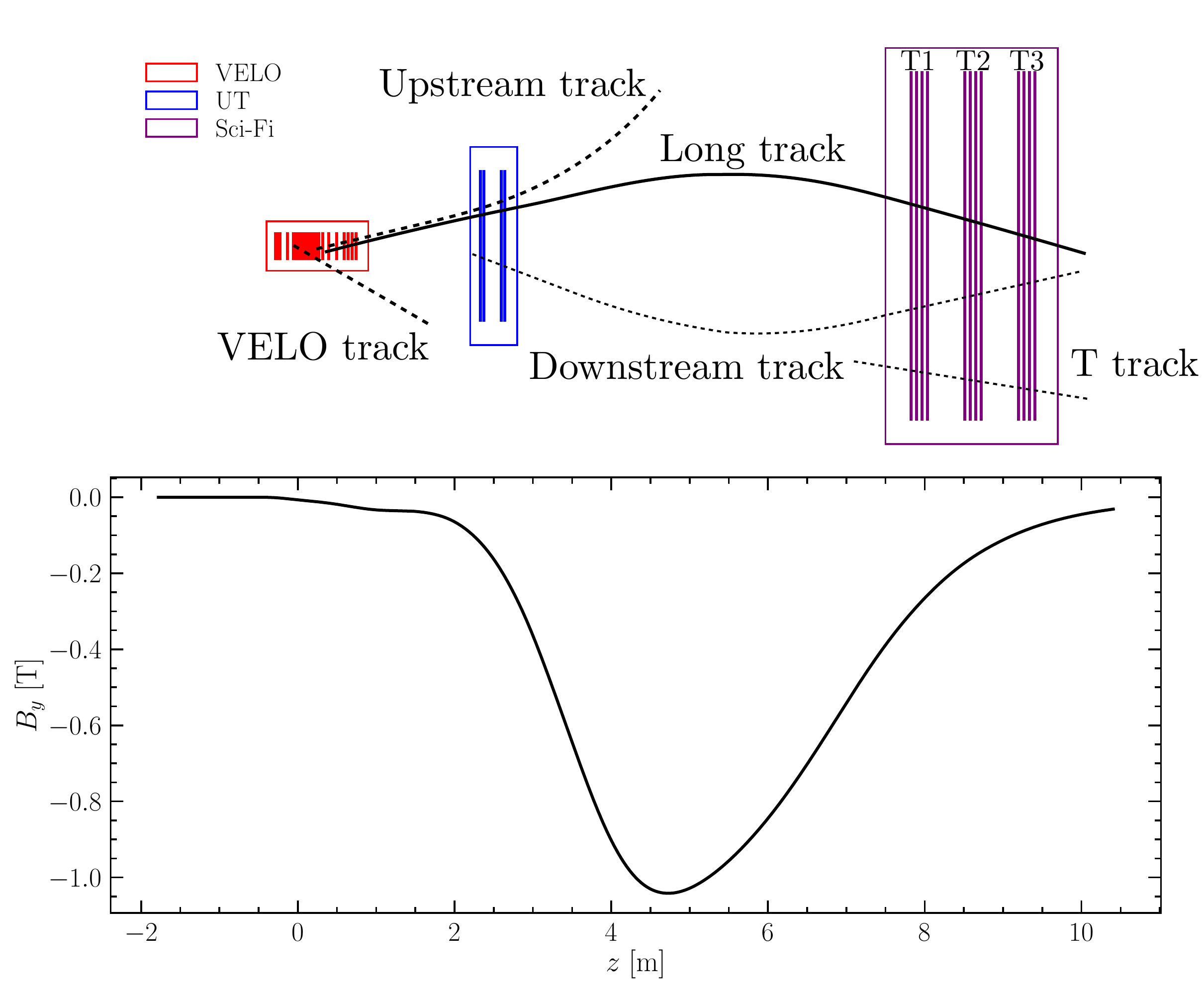}
    
    \caption{The geometric layout of LHCb's tracking system in the bending plane (\textbf{x-z}), shown in a right-handed coordinate system with the \textbf{z} 
    axis running from left to right. The top panel shows the different tracking detectors, described in the text, and the different track types reconstructed by LHCb's pattern recognition algorithms. The bottom panel shows the strength of the main component of LHCb's dipole magnetic field, orthogonal to the bending plane,
    as a function of the \textbf{z} position. The dipole magnet is 
    located between the UT and T1-T3 stations in this picture.}
    \label{fig:track_types}
\end{figure}

The Run~3 LHCb detector~\cite{LHCb:2023hlw} is a single-arm spectrometer optimised for the study of heavy flavour hadrons, primarily 
instrumented in the pseudorapidity range $2 < \eta < 5$. The layout of LHCb's tracking detectors and the corresponding dipole magnetic field intensity profile are shown in Figure~\ref{fig:track_types}. The tracking detectors consist of a silicon pixel vertex locator 
(VELO), located around the original proton-proton ($pp$) collisions region, the silicon-strip Upstream Tracker (UT), located upstream of the magnet, and the Scintillating Fibre tracker (FT), located downstream 
of the dipole magnet.
The tracks in the VELO are used to reconstruct the position of the $pp$ collisions as well as to discriminate between tracks originating in those collisions from tracks 
originating in the decays of other long-lived particles products. The UT and FT subdetectors are built from quadruplets of stations, in which each quadruplet is arranged in an 
\textbf{x-u-v-x} layout. The two \textbf{x} layers are oriented in the \textbf{x-y} plane with vertical silicon strips and fibres in UT and FT, respectively. The \textbf{u} and \textbf{v} layers are oriented in the \textbf{x-y} plane with strips (fibres) placed at $\pm 5$ 
degrees from vertical, which allows the three-dimensional position of the track to be reconstructed. The UT consists of 
one such quadruplet while the FT consists of three, referred as T1-T3 or collectively as the T-stations. In addition 
to tracking the LHCb detector is instrumented with particle identification subdetectors, which are not central to 
the algorithm presented in this paper.

During Run~3 the LHC will provide up to 30 million colliding bunch crossings (``events'') per second, each of which will 
contain an average of five individual $pp$ collisions in nominal LHCb $pp$ data-taking. This leads to an average of 
around 60 charged particles which traverse the VELO and the FT in each event.
The LHCb trigger system reduces the input data rate 
by a combination of selections, isolating the events containing processes of interest 
by computing high-level quantities of interest to physics analysis in real time.
This allows up to 90\% of data 
to be discarded even for selected bunch crossings~\cite{LHCb:2023hlw}. The majority of LHCb's physics analyses concern the decays of hadrons 
containing beauty or charm quarks. Because these hadrons have significant lifetimes, they typically travel around 1~cm 
in the laboratory frame before decaying. In addition, these hadrons produce decay products which have significant momentum 
transverse to the LHC beamline direction (\pt). It is this combination of displacement from the primary $pp$ collision \textbf{and} 
significant \pt which allows the bulk of LHCb's physics to be separated from generic Quantum Chromodynamic (QCD) processes occurring in LHC 
$pp$ collisions. Of the track types shown in Figure~\ref{fig:track_types} mainly long tracks allow both displacement and \pt 
to be precisely reconstructed, therefore, they form the backbone of both physics analyses and real-time selections at LHCb.

There are two strategies for reconstructing long tracks within the LHCb geometry. The first, known as ``forward''
tracking~\cite{Benayoun:2002hqa} , begins by reconstructing tracks in the VELO and subsequently extrapolates them through the UT and magnet region 
to the T-stations. The second, known as ``matching''~\cite{Gunther:2865000}, independently reconstructs tracks in the VELO and FT and 
subsequently matches them to each other, while the UT hits are used to improve track fit and help discriminating between correct and fake matches. The Looking 
Forward algorithm described in this paper follows the first strategy, but contains significant conceptual and practical 
modifications compared to previous LHCb forward algorithms ~\cite{Benayoun:2002hqa, Callot:2007mba} in order to better exploit highly parallel computing architectures.

The LHCb real-time tracking challenge in Run~3 can be put in a wider context by comparing it to the real-time tracking 
of the general-purpose LHC detectors: ATLAS and CMS. These aim to take data with around 60 $pp$ collisions 
per bunch crossing in Run~3~\cite{ATLAS:2023dns,CMS:2023gfb}, each of which produces an average of around 15 charged particles in the 
detector acceptance. However, their trigger systems are only allowing to reconstruct tracks in real-time at around 100~kHz, once interesting $pp$ 
collisions have been selected using information from calorimeters and muon detectors. These interesting collisions contain a 
greater than average number of tracks per event compared to LHCb~\cite{ATLAS:2021vfj,CMS-DP-2023-028}, nevertheless the overall number of tracks which the Run~3 ATLAS and CMS 
real-time systems have to reconstruct per second is around one order of magnitude smaller than it is for LHCb. The ALICE 
detector will reconstruct thousands~\cite{Antonioli:2013ppp} of charged particles in lead-lead collisions at 50~kHz in Run~3, again several 
factors smaller in terms of tracks per second than LHCb. The comparison of the number of reconstructed tracks per second for the four major experiments at the LHC is shown in Table~\ref{table:four_experiments}.  The real-time reconstruction of tracks in the LHCb detector 
is one of the biggest tracking challenges ever attempted in high-energy physics.

The future high-luminosity upgrades of ATLAS~\cite{ATLAS:2802799} and 
CMS~\cite{Collaboration:2759072} will operate real-time tracking at far higher rates and multiplicities, as will a planned future 
upgrade~\cite{LHCbCollaboration:2776420} of the LHCb detector. The study provided in this document will provide an insight for the future tracking challenges.

\begin{table}[h]
\begin{tabular}{l|llll}
                                                & LHCb               & ATLAS              & CMS                & ALICE              \\
                                                \hline
$\mathcal{L}$ {[}$cm^{-2} s^{-1}${]} & $2 \times 10^{33}$ & $2 \times 10^{34}$ & $2 \times 10^{34}$ & $6 \times 10^{27}$ \\
  pile-up              & 5                  & 60                 & 60                 & 1                  \\
reconstruction rate    & 30 MHz             & 100 kHz            & 100 kHz            & 50 kHz             \\
reconstructed tracks/s                          & 1800 M    & 90 M      & 90 M       & 10 M     
\end{tabular}
\caption{Comparison of the four major experiments at the LHC in terms of instantaneous luminosity $\mathcal{L}$, number of pp collisions per bunch crossing or pile-up, the rate at which the track reconstruction is performed and the number of reconstructed tracks per second.
pp collisions per bunch crossing}
\label{table:four_experiments}
\end{table}

\section{Requirements}

The first stage of LHCb's real-time processing is implemented~\cite{LHCbCollaboration:2717938} using 
A5000 NVIDIA GPUs hosted in up to 190 dual-socket servers with 32 physical cores and 512~GB of RAM per socket based on the 
AMD EPYC architecture. This processing pipeline is referred to as ``HLT1'' following standard LHCb nomenclature. 
One GPU was deployed per server during 2022 data-taking, increasing to two GPUs per server for 
2023 data-taking. The reconstruction algorithms are implemented in the Allen~\cite{Aaij:2019zbu} framework and run almost entirely 
on the GPU, with the host CPU(s) responsible for copying data to and from the GPU and a certain amount of auxillary 
monitoring tasks. During 2022, the LHC provided non-empty pp collisions at around 20 MHz which means each GPU must be able to process around 105~kHz of data input rate. In 2023, with a doubled number of GPUs and a LHC input rate of 30 MHz, each GPU can handle up to 80~kHz of data rate.

The track finding forms only one part of HLT1 employing around 50~\% of all available HLT1 resources. As we will show later, the Looking Forward algorithm can typically use around a 20~\% 
of the total resources for an HLT1 sequence which fits into the overall budget. 

The second stage of LHCb's real-time processing, ``HLT2'', is implemented using around 3500 dual-socket CPU servers~\cite{LHCb:2023hlw} 
of varying generations and core counts, primarily using Intel architectures. It is required to run at around 500~Hz on 
an ``average'' server representative of the overall data centre performance. LHCb's physics analyses use simulated events, 
which have been processed in the same way as data, to correct for detector inefficiencies. Simulation is processed 
exclusively using CPU servers. Therefore the Looking Forward algorithm is optimised for execution on GPUs but is 
required to compile for, and run on, CPU architectures. It must be sufficiently fast when executed on a CPU such that 
it can be deployed to simulate LHCb events without an increase of the overall cost of simulation production.

The algorithm is required (and tested) to be deterministic, however strict bitwise reproducibility when running in a 
different environment or on a different architecture is not a design requirement. Such reproducibility is impossible 
without emulators on parallel architectures in general, including between different CPU generations or instruction 
sets. However an emulator would clash with the earlier requirement that the Looking Forward algorithm remains computationally 
efficient when executed on a CPU. Strict reproducibility is not necessary because LHCb reconstructs its data only once, 
in real time. The selected data is annotated with provenance information that describes the objects which caused 
the real-time processing to keep it. Since LHCb simulation does not perfectly agree with data, the collaboration has 
a number of strategies~\cite{LHCb:2014nio,LHCb:2019gvd} for calibrating data-simulation differences. These methods can be applied to correct for 
small differences in the Looking Forward algorithm when executed on a CPU or on a GPU. The 
differences in the results of the Looking Forward algorithm when executed on a CPU or on a GPU must be at the permille level 
or smaller. The difference is considered negligible compared to other, typically percent level, data-simulation 
differences~\cite{LHCb:2014nio,LHCb:2019gvd} in LHCb's tracking.

In order to satisfy LHCb's physics requirements the Looking Forward algorithm is required to have an efficiency which 
is as close as possible to the forward tracking which runs, under much more relaxed computational constraints, in HLT2. 
In particular, performance for tracks with $\pt > 500$~MeV and tracks produced in 
the decays of beauty hadrons are required to be within a few percent different from what is achieved in HLT2. 
The algorithm is required to be configurable such that physics performance can be smoothly traded off against computational 
complexity, and to be robust against inefficiencies in the detector itself. As the UT subdetector was not installed in 
time for the 2022 data-taking, the changes to the Looking Forward algorithm to deal with its absence, which are documented 
in this paper, represent a stress test of the robustness requirement.

\section{Benchmarking setup}

High energy physics experiments have traditionally benchmarked algorithmic performance in terms of wall clock time, i.e. 
how many seconds a given algorithm takes to process an average bunch crossing on a reference computing node. This metric 
can be suitable for serial algorithms, but is inherently flawed when 
benchmarking highly parallel architectures. In such a regime 
the correlation between a given algorithm's reported resource usage and 
the overall sequence throughput is weak at best. For these reasons our primary computational benchmarking metric is the overall 
throughput of the nominal LHCb HLT1 sequence. We additionally cite the GPU resource usage of Looking Forward, as reported by 
the NVIDIA Nsight Compute profiling tool, and compare it to other parts of the HLT1 sequence.

Computational benchmarking is carried out using a dedicated testbench server hosting a range of NVIDIA GPU cards, as well 
as a dual-socket server equipped with AMD EPYC 7502 CPUs. 
The specifics of the hardware are detailed in Table~\ref{table:benchmark}.

\begin{table}[t]
    \centering
    \begin{tabular}{l||l|l|l|l|l}
    \hline
    Unit & \# cores & Max freq.  & Cache  & DRAM & TDP  \\
    &  &  [GHz] &  [MiB - L2] &  [GiB] &  [W] \\
    \hline 
    GeForce  & \multirow{2}{*}{10496} &  \multirow{2}{*}{1.69} &  \multirow{2}{*}{6} &  24 &  \multirow{2}{*}{350} \\
    RTX 3090  &  &  &   & GDDR6X   &  \\
    \hline
    RTX A5000 & \multirow{2}{*}{8192} &  \multirow{2}{*}{1.69} &  \multirow{2}{*}{6} & 24 &  \multirow{2}{*}{230} \\
    &  &  &   & GDDR6  &  \\
    \hline
    GeForce & \multirow{2}{*}{4352} &  \multirow{2}{*}{1.54} &  \multirow{2}{*}{6} &  11 &  \multirow{2}{*}{250} \\
    RTX 2080 Ti &  &   &   & GDDR6   &   \\
    \hline
    EPYC 7502 & \multirow{2}{*}{32} &  \multirow{2}{*}{3.35} &  \multirow{2}{*}{128 (L3)} & 64 &  \multirow{2}{*}{180} \\
    32-Core &  &   &   & DDR4   &   \\
    \hline
    
    \end{tabular}
    \caption{GPU and CPU hardware used to benchmark the HLT1 throughput. A comparison is presented for three NVIDIA graphic cards (GeForce RTX 3090, RTX A5000 and GeForce  RTX 2080 Ti) and an AMD EPYC 7502 CPU. The number of cores, the maximum frequency, the cache, the dynamic random-access memory (DRAM) and the thermal design power (TDP) are shown.} 
    \label{table:benchmark}
    \end{table}

Throughput is 
measured on samples of simulated ``minimum bias'' events produced with the full LHCb detector simulation under nominal 
Run~3 conditions. Throughput is benchmarked as a function of the number of $pp$ collisions in the event to study the scalability of the Looking Forward algorithm in different running conditions.

Physics benchmarking is carried out using full LHCb detector simulation under Run~3 conditions. The basic performance 
metrics are described in Ref.~\cite{Li:2752971} and recapitulated here for convenience. Reconstructed tracks are matched to ground 
truth information in the simulated samples to determine whether they represent a genuine charged particle trajectory. It is 
required that more than 70\% of detector hits on a reconstructed track and a ground truth particle match in order to consider 
that particle correctly reconstructed. The algorithm's efficiency to correctly reconstruct particles is measured with respect 
to ``reconstructible'' charged particles which leave a minimal number of hits in the tracking detectors (VELO, UT, FT). The 
fake rate is defined as the fraction of reconstructed tracks which are not matched to a ground truth particle. The efficiency 
and fake rate are benchmarked as a function of kinematic and geometric properties of the particles. The resolution on the 
reconstructed track momenta and the resolution on the track slopes are also 
reported. The performance of the algorithm for CPU and GPU architectures is compared by processing the same set of simulated 
events on each architecture and comparing the reconstructed quantities of interest on a track-by-track basis between the two. 

\section{Propagation of tracks in LHCb's magnetic field}

In order to accurately reconstruct charged particle trajectories it is essential to have an accurate model of their 
bending in the experiment's magnetic field. The general equation of motion of a charged particle with momentum $\vec{p}$, 
charge $q$ and velocity $\vec{v}$ in a magnetic field $\vec{B}$ is
$$ \frac{\text{d}\vec{p}}{\text{d}t} = q \vec{v} \times \vec{B},$$
which leads to the following equations in the three dimensions considering the momentum components $p_x$, $p_y$, and $p_z$:
$$ \frac{\text{d}p_x}{\text{d}z} = q (t_y B_z - B_y),$$
$$ \frac{\text{d}p_y}{\text{d}z} = q (B_x - t_x B_z),$$
$$ \frac{\text{d}p_z}{\text{d}z} = q (t_x B_y - t_y B_x).$$

Here $t_x = p_x/p_z = \frac{\text{d}x}{\text{d}z}$ and $t_y = p_y/p_z = \frac{\text{d}y}{\text{d}z}$ are the track slopes.
 
The two differential equation for the tracks slopes in the \textbf{x-z} and \textbf{y-z} planes respectively are
$$  \frac{\text{d}t_x}{\text{d}z}  = \frac{q}{p} \sqrt{1 + t_x^2 + t_y^2} (t_x t_y B_x - (1 + t_x^2) B_y + t_y B_z),$$
$$  \frac{\text{d}t_y}{\text{d}z}  = \frac{q}{p} \sqrt{1 + t_x^2 + t_y^2} ( (1 + t_y^2) B_x -  t_x t_y B_y - t_x B_z).$$

The LHCb magnetic field has been mapped~\cite{1018421,1324843} in a series of dedicated measurement campaigns. Its dominant component is $B_y$, 
so that tracks traversing the magnet are deviated almost entirely in \textbf{x}, with deviations in \textbf{y} being 
smaller than the detector resolution in most cases. 
Assuming therefore that $B_x \sim 0$, $B_z \sim 0$  then for small $|t_x|$ and $|t_y|$ the earlier equations can be approximated 
keeping only the first order terms, leading to
$$ \frac{\text{d}^2x}{\text{d}z^2} = \frac{t_x}{z} \sim - \frac{q}{p} B_y,$$
$$ \frac{\text{d}^2y}{\text{d}z^2} = \frac{t_y}{z} \sim 0.$$
The y-z equation results in a simple linear model,
$$ y(z) = y_0 + t_y (z - z_0),$$
where $y_0$ is the $y$ coordinate at a given reference position $z_0$.

For the \textbf{x-z} track projection, the $B_y$ dependence on $z$ can be parameterised at first order as
$$ B_y(z) \sim B_0 + B_1 (z - z_0)$$ in the magnetic field tails within the FT acceptance, assuming a linear decrease of the $B_{y}$ component along the $z$ direction.
This in turn leads to the following dependence of the track's \textbf{x} position as a function of \textbf{z}, within the FT acceptance
\begin{equation}
\label{eq:trackfitmodel}
    \begin{split}
        x(z)&= x_0 + t_x(z - z_0) +  \\
            &  \frac{q}{2p}B_0 (z - z_0)^2 (1 + \text{dRatio}(z - z_0)),
    \end{split}
\end{equation}
where $x_0$ is the coordinate at a reference position $z_0$ and the quantity dRatio = $\frac{B_1}{3B_0}$ is roughly constant 
in the region where the tracks are extrapolated.
This parameterization avoids the slowdowns due to real-time usage of the LHCb magnetic field maps, such as memory access, copying to GPU, GPU memory size consumption, ... by keeping the precision required by the HLT1 reconstruction.

The track fit model within the FT acceptance region depends linearly on five adjustable parameters: two related to the \textbf{y-z} projection, 
$y_0$ and $t_y$, and three related to the \textbf{x-z} projection, $x_0$, $t_x$ and $B_0\frac{q}{p}$, similarly as what is in ~\cite{Aiola:2020ydy}.

\section{Algorithm logic}

The Looking Forward algorithm begins with tracks which have been reconstructed by the search by triplet~\cite{CamporaPerez:2021jhc} algorithm 
in the VELO and the CompassUT~\cite{FernandezDeclara:2019ycx} algorithm in the UT. The track state used to define the parameters in Equation~\eqref{eq:trackfitmodel} is calculated at the downstream end of the UT detector. 

The Looking Forward algorithm is composed of four main steps:
\begin{enumerate}
    \item \textbf{Defining the search windows opening}: search windows tolerances are evaluated in each SciFi layer extrapolating the input VELO-UT tracks to reduce the number of hit combinations;
    \item  \textbf{Triplet seeding}: triplet of hits are combined using the x-layers information;
    \item  \textbf{Extending triplets to other layers}: triplets are extended to the other layers and a track fit is performed;
    \item  \textbf{Quality filter}: candidates are filtered and selected thanks to a track quality factor based on the fit $\chi^2$.
\end{enumerate} 
After these four steps, the calculation of the momentum of the particle and evaluation of the track states are performed.

\subsection{Defining the search windows}

Defining the search windows in the FT is a necessary step to reduce the number of hit combinations. Each SciFi layer contains an average number of hits of $n_{hits} \sim 400$ which are combined into triplets, reaching a maximum number of combinatorics of $n_{hits} \times n_{hits} \times n_{hits}$ multiplied by the number of input tracks. In order to reduce the algorithm complexity, search windows are defined in each SciFi layer using the input track information.

The track's slope in the non-bending plane is assumed constant, $t_y^{\textrm{Velo}} = t_y^{\textrm{FT}}$. 
The $t_y$ information defines whether the particle is traveling in upper or lower half of the FT, since 
tracks emerging from the VELO are unlikely to cross both halves of the FT. This allows one half of the FT hits to be 
removed from consideration, immediately reducing the computational burden.

\begin{figure}[t]
    \centering
    \includegraphics[width=0.45\textwidth]{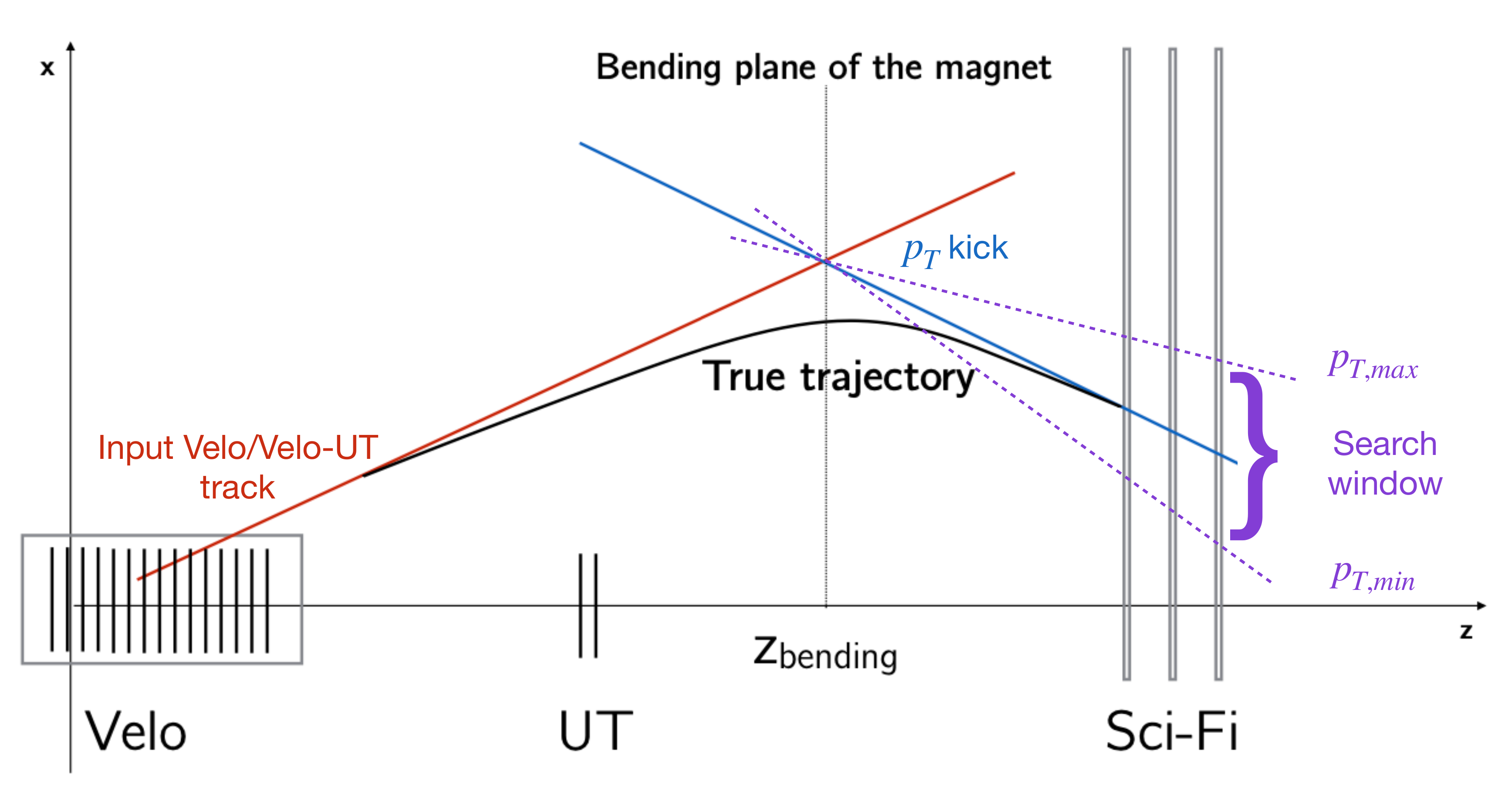}
    \caption{ An illustration of the \pt-kick method \cite{Quagliani:2296404}.}
    \label{fig:pt_kick}
\end{figure}

A fast momentum estimation is the so-called \pt-kick method \cite{Bowen:1635665,Quagliani:2296404}.
The effect of the magnetic field between two detectors is parametrised as an instantaneous kick to the 
momentum vector at the center of the magnet, $z_{bending}$, as illustrated in Figure~\ref{fig:pt_kick}.
The momentum kick, $\Delta \vec{p}$ is defined as
$$ \Delta \vec{p}  = q \cdot \int d \vec{l} \times \vec{B},$$
with an integral of $\vec{B}$ the magnetic field along the path followed by the track.

Since, as discussed earlier, the bending can be assumed to occur only in the \textbf{x} plane with $B_x \sim 0$, $B_z \sim 0$  and for small $|t_x|$ and $|t_y|$, this simplifies to
\begin{equation}
\begin{split}
\label{eq:pt_kick}
    \frac{q}{p}&= \frac{1}{q \cdot \int |d \vec{l} \times \vec{B}|_x }\\ 
    & \times \Bigg(\frac{t_{x,f}}{\sqrt{1 + t_{x,f}^2 + t_{y,f}^2}} - \frac{t_{x,i}}{\sqrt{1 + t_{x,i}^2 + t_{y,i}^2}}\Bigg),
\end{split}
\end{equation}
with $t_{x,i}$ ($t_{x,f}$) and $t_{y,i}$ ($t_{y,f}$) respectively the x and y initial (final) slopes of the track's segments.
In this way, it is sufficient to know the initial and final track slopes to have an estimate of the particle's momentum.
Alternatively, given a VELO-UT track and an assumed momentum, the \pt-kick method can be used to define search windows 
in the FT and further reduce the number of hits considered for a specific input track.
The values of $ \frac{1}{q \cdot \int |d \vec{l} \times \vec{B}|_x }$ are stored in a lookup table and used by the Looking 
Forward algorithm for fast processing. 

The extrapolated center of the search windows for each VELO-UT track is defined in every FT layer $i$ as
$x_{\textrm{extrap}}^i$, $y_{\textrm{extrap}}^i$,
and is calculated by extrapolating the track from the \textbf{z}-position of the center of the magnet 
with the \pt-kick method. The size of the search windows in the x layers is defined as:
$$ x_{\textrm{min}} = x_{\textrm{extrap}} - x_{\textrm{tol}} - x_{\textrm{asym}}^{\textrm{left}}$$
$$ x_{\textrm{max}} = x_{\textrm{extrap}} + x_{\textrm{tol}} + x_{\textrm{asym}}^{\textrm{right}}$$
with $[x_{\textrm{min}},x_{\textrm{max}}]$ the search window region in x, $x_{\textrm{tol}}$ the window size or tolerance factor,
and $x_{\textrm{asym}}^{\textrm{left/right}}$ a magnetic field asymmetry factor.
The $x_{\textrm{tol}}$ factor defines the search window size as:
$$ x_{\textrm{tol}} = 150~{\mathrm{mm}} + \frac{2000 ~\textrm{GeV/c}}{p} $$
as a function of the particle's momentum $p$.
The last term, $x_{\textrm{asym}}^{\textrm{left/right}}$, takes into account the fact that the magnetic field is not a pure dipole, but is affected by fringe effects mainly away from its center.
It is defined as:
\begin{equation}
    \begin{split}
    x_{\text{asym}}^{\text{left}} &= 
    \begin{cases}
        100~{\mathrm{mm}}, & \text{if } q=+1 \text{ } \& \text{ MU OR } q=-1 \text{ }\& \text{ MD}\\
        0~{\mathrm{mm}} ,  & \text{otherwise}
    \end{cases}
    \\
    x_{\text{asym}}^{\text{right}} &=
    \begin{cases}
        100~{\mathrm{mm}}, & \text{if } q=-1 \text{ } \& \text{ MU OR } q=+1 \text{ }\& \text{ MD}\\
        0~{\mathrm{mm}} ,  & \text{otherwise}
    \end{cases}
    \end{split}
\end{equation}
depending on the charge of the particle $q$ and the polarity of the magnet (MU or MD).

The center of the search window in the \textbf{u} and \textbf{v} layers is defined by extrapolating the track 
as a straight line from the neighbouring \textbf{x} layer. This extrapolation is corrected by a 
$\pm \sin{5^{\circ}}$ factor of the \textbf{u} and \textbf{v} layer, while a tolerance of $\pm 800 ~{\mathrm{mm}}$ around 
the center defines the search window for the hits. 

The maximum number of hits allowed in the search window for any single layer is defined to be $n_{\mathrm{hits}}^{\mathrm{window}}$, whose values depending on the sequence is reported in Table~\ref{table:tolerances}. This limit 
constraints the number of combinations passed to the next stage and consequently the memory consumption of 
the algorithm. If there are more than $n_{\mathrm{hits}}^{\mathrm{window}}$ hits in a given search window, the hits kept are chosen 
symmetrically around the center of the window.

\subsection{Triplet Seeding}

Once the search windows have been defined, the pattern recognition begins by forming triplets of hits ($h_0$,$h_1$,$h_2$) 
from the \textbf{x} layers. Two configurations of layers are used for this search: either the first layer of each T-station 
or the last layer of each T-station. First, all possible hit doublets are formed for each VELO-UT track from the first and 
last station ($h_0$,$h_2$). Two conditions are used to filter doublet candidates:
\begin{enumerate}
    \item If the VELO-UT input track has momentum $<$ 5 \gevc, the bending of track from the VELO to the UT is required to be 
    in the same direction as the bending of the track from the VELO to the FT doublet;
    \item A maximal tolerance in the \textbf{x}-opening at the \textbf{z} position of the magnet $z_{\text{bending}}$ is defined 
    as a function of the momentum $p^{{\mathrm{VeloUT}}}$ of the VELO-UT track and on the doublet's slope $t_x^{\text{doublet}}$. It 
    varies from 8 to 40$~{\mathrm{mm}}$. The difference in the straight line extrapolation of the VELO-UT track and the FT doublet to 
    the center of the magnet is required to be smaller than this tolerance.
\end{enumerate}
The filtered doublets are upgraded to triplets by adding the hit in the corresponding second T-station using a straight 
line extrapolation corrected by
$$ x_1^{\text{expected}} = z_1 \cdot t_x^{(h_0\text{-}h_2)} + (x_0 - t_x^{(h_0\text{-}h_2)} \cdot z_0) \cdot K_{x_1},$$
where $x_1^{\text{expected}}$ is the extrapolated \textbf{x}-position of the $h_1$ hit, $z_i$ and $x_i$ are the \textbf{z}- 
and \textbf{x}-positions of the respective hits, $t_x^{(h_0\text{-}h_2)}$ is the slope in the bending plane between the 
$h_0$ and $h_2$ hits and $K_{x_1}$ is a sagitta-like correction factor.

Because of the residual magnetic field in the T-stations the value of $K_{x_1}$ is different depending on whether the 
first or last T-station layers are being used:
$K^{\mathrm{first}}_{x_1} = 1.00177513$ and $K^{\mathrm{last}}_{x_1} = 1.00142634$.

The hit with \textbf{x}-value closest to $x_1^{\mathrm{expected}}$ is added to form the triplet candidate 
if the $ \chi^2_{\mathrm{triplet}} = (x_1^{\mathrm{hit}} - x_1^{\mathrm{expected}})^2 < T_{\mathrm{triplet}}$, where 
$T_{\mathrm{triplet}}$ is a tolerance reported in Table~\ref{table:tolerances}.
The maximum number of selected triplets is a configurable parameter. A parameter scan shows that the optimal number 
in current LHCb data-taking conditions is $n_{\mathrm{triplets}}^{\mathrm{track}} = 12$ triplets per VeloUT track.
If the maximum number is exceeded, the 12 candidates with the smallest $\chi^2_{\text{triplet}}$ are selected.
     
\subsection{Extending triplets to other layers}

Triplets are extended to the remaining FT layers in order to form full track candidates. A track candidate must 
contain at least 9 hits, with at least one hit in the \textbf{u} or \textbf{v} layers per T-station. This threshold 
is chosen to keep the fake rate manageable.
 
The extrapolation is performed using Equation~\ref{eq:trackfitmodel}.
Tracklets are first extended to the three missing \textbf{x} layers, computing the expected \textbf{x}-position of the 
hit in layer $i$:
\begin{equation}
x^{\textrm{expected}}_i(z) = a_x + t_x \cdot dz_i + c_x \cdot dz_i^2 \cdot (1 + \textrm{dRatio} \cdot dz_i)
\label{eq:dratio_model_x_expected}
\end{equation}
where the values of $a_x$,$t_x$,$c_x$ and dRatio are evaluated using the position information from the triplet hits. 
The effect of the magnetic field is parametarized~\cite{Quagliani:2296404} as a function of $dz_i = z_i - z^{\text{ref}}$, 
the difference between the \textbf{z}-position of the $i$ layer and a reference plane at $z^{\text{ref}} = 8520 ~{\mathrm{mm}}$.
For each \textbf{x}-layer $i$, the hit with \textbf{x}-value closest to $x^{\textrm{expected}}$ is added to the 
triplet if $ \chi^2_{x_i} = (x_i^{\text{hit}} - x_i^{\text{expected}})^2 < T_x$, where the  
$T_x$ tolerance factors are reported in Table~\ref{table:tolerances}.

Each tracklet candidate is then extended looking for hits in the \textbf{u} and \textbf{v} layers,
where Equation~\ref{eq:dratio_model_x_expected} is corrected for the $\pm \sin{5^{\circ}}$ angle of those layers.
Hits are again added according to a $\chi^2$ tolerance listed in Table~\ref{table:tolerances}.
However for the \textbf{u} and \textbf{v} layers the tolerance is a function of the track's slopes in 
order to allow more generous windows for more peripheral tracks in both \textbf{x} and \textbf{y}.

\begin{table}[t]
    \centering
    \begin{tabular}{l|ll}
    \hline
    parameters & Forward & Forward no-UT\\
    \hline
    $T_{\text{triplet}}$ &  8.0 & 2.0\\
    $T_x$ &  2.0 & 0.5\\
    $T_{uv}$ &  50 $\cdot (t_x + t_y)$ & 15 $\cdot (t_x + t_y)$\\
    $T_{t_y}$ &   0.02 & 0.003 \\
    $T_{y}$ &   800 & 800 \\
    $T_Q$ &  0.5 & 0.5 \\
    \hline
    $n_{\mathrm{hits}}^{\mathrm{window}}$ & 32 & 64 \\
    $n_{triplets}^{track}$ & 12 & 20 \\
    \hline
    
    \end{tabular}
    \caption{Parameters and tolerances used in the forward reconstruction.
    The tolerance relative to the \textbf{y}-position $T_{y}$ has a much larger value compared to 
    the \textbf{x}-position due to the fact that the information only comes from the \textbf{u} and \textbf{v} layers 
    and therefore has a poorer resolution. $T_{t_y}$ represents the \textbf{y}-\textbf{z} slope difference tolerance 
    when performing the mean squared fit in the \textbf{y} direction.
    The maximum number of hits per search window and triplets considered for the same input tracks are also reported.        }
    \label{table:tolerances}
    \end{table}

\subsection{Quality Filter}

The reconstructed tracks are filtered in order to reduce the fake rate. 
First, a linear least square fit in the y-direction is performed on the candidates 
evaluating the $t_y^{\text{expected}}$ slope and $\chi^2$ value $ \chi^2_{y_i} =  (y_i^{\text{hit}} - y_i^{\text{expected}})^2$ based on the expected and measured \textbf{y}-positions of the \textbf{u} and \textbf{v} hits.
The difference in the measured and expected slopes in the non-bending plane is required to be less than a 
tolerance $T_{t_y}$ whose value is given in Table~\ref{table:tolerances}.

The total \textbf{x} and \textbf{y} quality factors are determined as:
\begin{equation}
    \begin{split}
    Q_x &= \sum_{i=0}^{N_{\textrm{x hits}}} \frac{\chi^2_{x_i}}{T_x} + \sum_{j=0}^{N_{\textrm{uv hits}}} \frac{\chi^2_{uv_j}}{T_{uv}}, \\
    Q_y &= \sum_{i=0}^{N_{\textrm{uv hits}}} \frac{\chi^2_{y_i}}{T_y}.
    \end{split}
\end{equation}
Here  $\chi^2_{x_i}$ and $\chi^2_{uv_j}$ are the $\chi^2$ values evaluated previously for each respective hit 
normalized by their tolerances $T_x$ and $T_{uv}$, while 
$\chi^2_{\text{y-fit}}$ is the \textbf{y}-fit $\chi^2$ normalized to its tolerance $T_y$. 

Each track is assigned an overall quality factor
\begin{equation}
    Q = (\frac{Q_x}{nDoF_x} +  \frac{Q_y}{nDoF_y}) \cdot C(n_{ {\mathrm{hits}} } ),
\end{equation}
where $nDoF_{x\text{-}y}$ is the number of fit degrees of freedom in the \textbf{x} and \textbf{y} directions and 
$C(n_{\text{hits}})$ is a multiplicative parameter dependent on the number of hits on the track candidate.
The values of $C(n_{\mathrm{hits}})$ are reported in Table~\ref{table:C_nhits} and favor tracks made out of a 
greater number of hits. The number of degrees of freedom $nDoF_x$ when performing the fit in the \textbf{x}-\textbf{z} 
plane is $nDoF_x = n_{\mathrm{hits}}^{\mathrm{total}} - 3$ as only information from the three \textbf{x}-layers is used.
In the \textbf{y}-\textbf{z} plane, $nDoF_y = n_{\mathrm{hits}}^{uv} - 2$ as only two uv-hits are used for the linear fit.

Track candidates are accepted as reconstructed if their quality factor is lower than the tolerance $T_Q$
given in Table~\ref{table:tolerances}. If more than one candidate is found for a given VELO-UT track, 
the one with lowest value of $T_Q$ is kept.

\begin{table}[t]
    \centering
    \begin{tabular}{l|l}
    \hline
    $n_{\text{hits}}$ & $C(n_{\text{hits}})$ value    \\
    \hline
    9 &  5.0 \\
    10 &  1.0\\
    11 &   0.8\\
    12 &  0.5 \\
    \hline
    
    \end{tabular}
    \caption{Values of the multiplicative parameter $C(n_{\mathrm{hits}})$ as a function of the number of hits on the 
    track candidate.}
    \label{table:C_nhits}
\end{table}



\subsection{No-UT tuning}

The fact that the UT tracker was not installed in time for the 2022 data-taking required the forward tracking to be retuned 
to extrapolate tracks directly from the VELO to the FT. Since VELO tracks do not have a charge or momentum estimate, their 
search windows must be double-sided and significantly wider than those of the VELO-UT tracks.

\begin{figure}[t]
    \centering
    \includegraphics[width=0.48\textwidth]{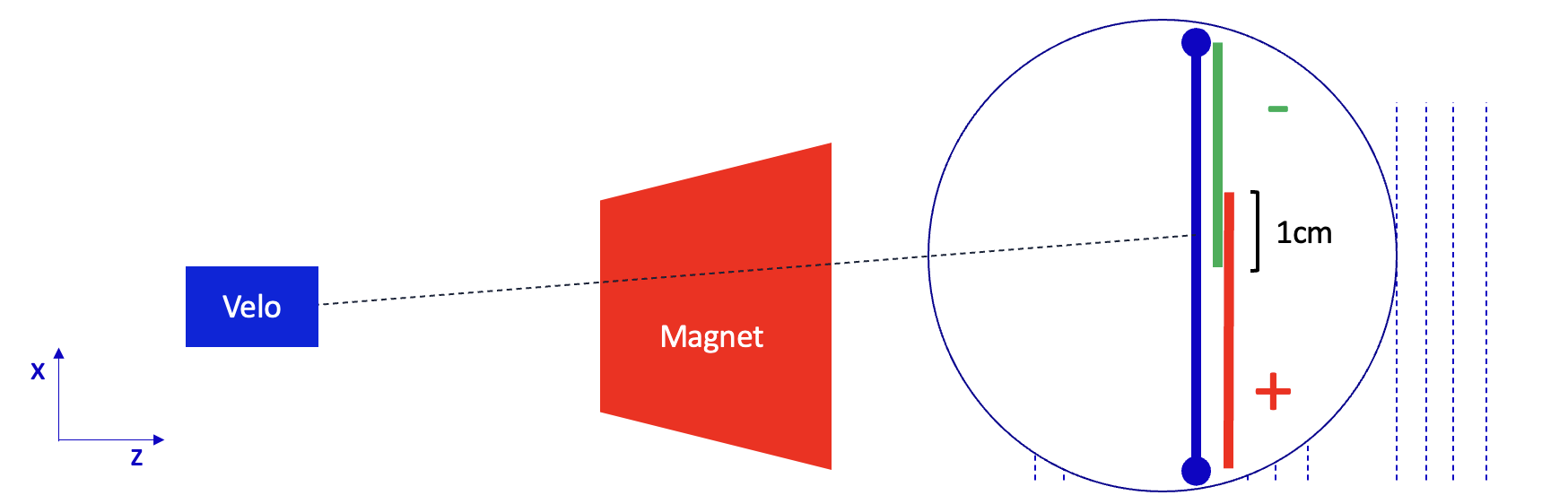}

    \caption{Scheme of the search window strategy for the forward tracking in the absence of the UT.
    A small overlap of 1~cm is allowed between the two windows to account for very high momentum tracks, 
    which are expected to travel in a straight line.}
    \label{fig:sw_no_ut_scheme}
\end{figure}

The VELO input tracks are extrapolated as a straight line directly to the various FT layers.
As the tracks momenta are unknown, they are set to a minimum value (maximum allowed curvature) which is a tunable 
parameter of the algorithm. The baseline no-UT thresholds are either a momentum of 5~GeV or a transverse momentum 
of 1~GeV. This transverse momentum threshold is converted into a momentum threshold using the VELO track slopes, and 
the tighter of the two momentum thresholds is used to determine the search window.
The VELO track $t_y$ slope is still used to determine which half of the FT, lower or upper, is used when 
extending the track to the various layers. Figure~\ref{fig:sw_no_ut_scheme} shows a scheme of the search window 
strategy.

The tolerance of the search window is evaluated with the following polynomial approximation~\cite{Hasse:2706588}
\begin{equation}
\begin{split}
\raggedleft 
x_{tol} &=  \frac{1}{p^{\mathrm{Velo}}}\Bigg[ c_0 + t_x (c_1 + c_2 t_x) \\ 
 &+ t_y^2 (c_3 + t_x (c_4 + c_5 t_x)) \\
 &+ \frac{1}{p^{\mathrm{Velo}}} (c_6 t_x + c_7 \frac{1}{p^{\mathrm{Velo}}})\Bigg]
\end{split}
\label{eq:tol_noUT}
\end{equation}

where $p^{\mathrm{Velo}}$ is the assumed VELO track momentum, $c_i$ are the coefficients defined in Table~\ref{table:poly_tol},
and $t_x$ and $t_y$ are the slopes in the \textbf{x}-\textbf{z} and \textbf{y}-\textbf{z} plane of the VELO input track.
The polynomial approximation is used to determine a curvature correction assuming a minimum momentum.
It is applied instead of the more common Runge-Kutta method~\cite{Kutta}, which provides a higher precision  
but is too computationally demanding for our requirements.

\begin{table}[t]
    \centering
    \begin{tabular}{llll}
    \hline
    $c_0$ & $c_1$ & $c_2$ & $c_3$  \\
    \hline
    4824.3 & 426.3 & 7071.1 & 12080.4 \\
    \hline
    $c_4$ & $c_5$ & $c_6$ & $c_7$ \\
    \hline
    14077.8 & 13909.3 & 9315.3 & 3209.5 \\
    \hline
    
    \end{tabular}
    \caption{Polynomial coefficients tuned on simulation~\cite{Hasse:2706588} approximating 
    the curvature of a VELO track with an assumed minimum momentum.}
    \label{table:poly_tol}
\end{table}

The double-sided \textbf{x}-layer search windows are defined as
\begin{equation}
    \begin{split}
x_{\textrm{min}}^{LEFT}\,\,\, &= x_{\textrm{extrap}} - x_{\textrm{tol}}; \\
x_{\textrm{max}}^{LEFT}\,\,\, &= x_{\textrm{extrap}} + x_{\textrm{overlap}}; \\
x_{\textrm{min}}^{RIGHT}      &= x_{\textrm{extrap}} - x_{\textrm{overlap}}; \\
x_{\textrm{max}}^{RIGHT}      &= x_{\textrm{extrap}} + x_{\textrm{tol}}. \\
    \end{split}
\label{eq:noUT_size_window}
\end{equation}
Here $x_{\textrm{extrap}}$ is the VELO track's straight line extrapolation \textbf{x}-position in the layer, 
$x_{\textrm{tol}}$ the window tolerance defined in Equation~\ref{eq:tol_noUT},
and $x_{\textrm{overlap}} = 5~{\mathrm{mm}}$ is an overlap factor between the two windows.
The overlap is necessary in order to ensure that hits on very high momentum tracks are considered for both charge 
assumptions, minimizing inefficiencies in these cases.

The \textbf{u} and \textbf{v} layer search windows are computed analogously to Equation~\ref{eq:noUT_size_window}, with 
a tolerance of $x_{\mathrm{tol}} = 1200~{\mathrm{mm}}$. The extrapolation to these layers is similarly analogous to the case where 
UT information is used.

The tunable parameters reported in Table~\ref{table:tolerances} are optimised for no-UT configuration.
The limit of hits for each search window is set to $n_{\mathrm{hits}}^{\mathrm{window}} = 64$  hits, symmetrically around the center of each window.
It is double that of the forward tracking with the UT because of the much larger search windows. In order to compensate the higher number of combinations, tighter $\chi^2$ thresholds are defined in the next section to handle the fake rate. 


The triplet seeding is analogous to the algorithm which uses UT information, with the combinations performed separately for each charge assumption. The \textbf{x} tolerance at the \textbf{z} position of the magnet is fixed to 10~mm in the 
baseline no-UT tuning since the algorithm is primarily searching for higher momentum tracks which bend less.
The maximum number of selected triplets is increased to $n_{\mathrm{triplets}}^{\mathrm{track}} = 20$ because of the wider search windows. From this point on 
the algorithm follows the same steps as for the case with UT information, with generally tighter $\chi^2$ tolerance
requirements which improve the computational performance and reduce the fake rate.
The charge and the momentum evaluations are performed with the \pt-kick method, as explained in the following.

\subsection{Calculation of momentum and track states}
The evaluation of the track momentum is obtained using the Equation \eqref{eq:pt_kick}.  Here the term associated to the integrated magnetic field along the track trajectory is parameterised
 according to a fourth order polynomial expansion as a function of $\mathrm{dSlope} = t_{x}^{\mathrm{FT}}-t_{x}^{\mathrm{Velo}}$, the measured variation of the track slope in the bending plane.
 The value of $\int{ \overrightarrow{B}\times d\overrightarrow{L}}$ is evaluated with a dedicated parameterisation of fourth order polynomial expansion
 $\int{ \overrightarrow{B}\times d\overrightarrow{L}} = F(t_{x}^{\mathrm{Velo}}, t_{y}^{\mathrm{Velo}}, \mathrm{dSlope})$, where the coefficients of the polynomial function $F$depend on 
a given track's entry slope direction in the field:
\begin{equation}
\begin{split}
F(t_{x}^{\mathrm{Velo}}, t_{y}^{\mathrm{Velo}} , \mathrm{dSlope})   
            &=\sum_{i=0}^{4} c_{i}\mathrm{dSlope}^{i}
\end{split}
\end{equation}

\begin{figure}
    \centering
    \includegraphics[width=0.45\textwidth]{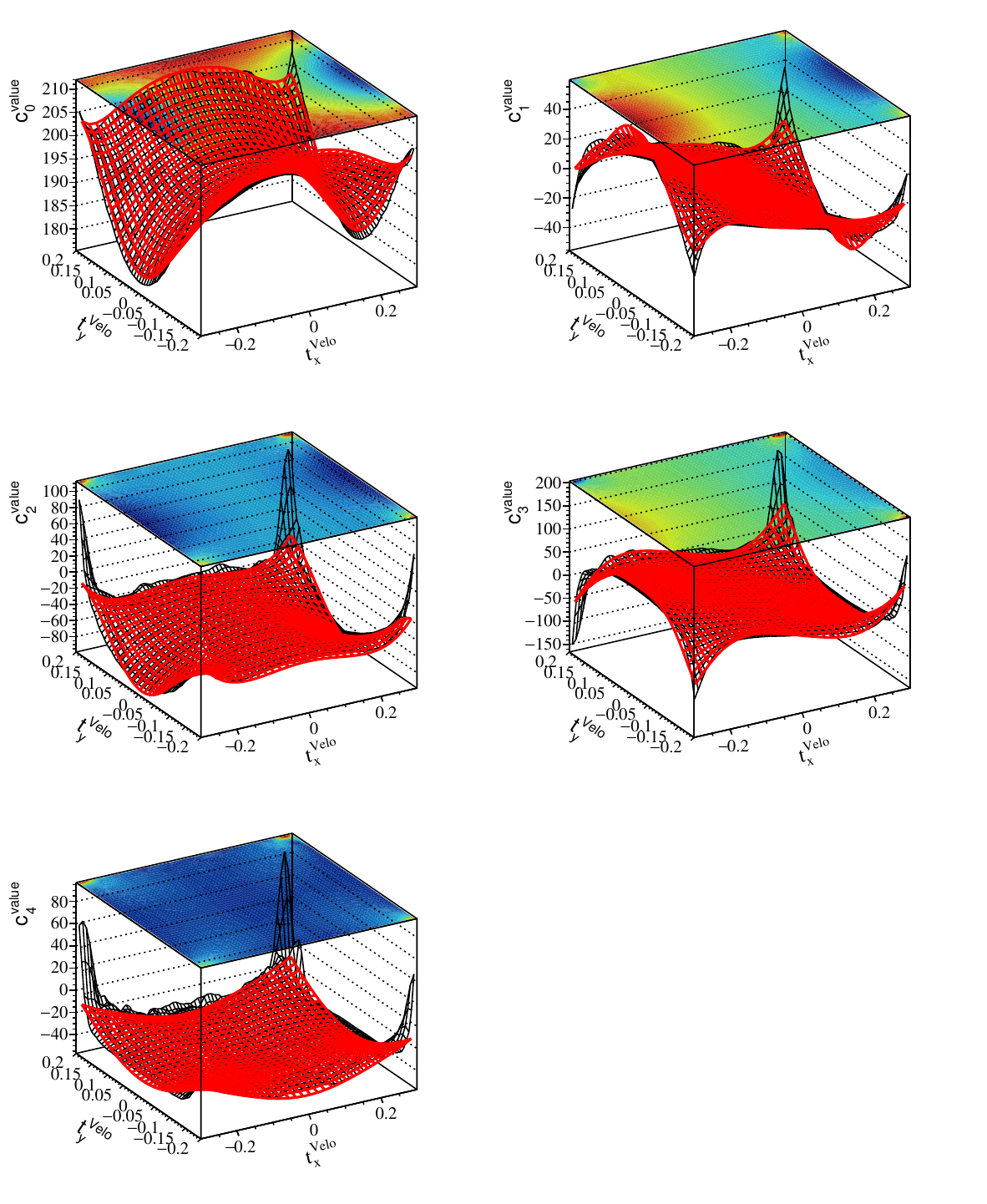}
    \caption{Fits determining the fourth order expansion polynomial terms for $c_{i}$. The $c_{i}$ values from the polynomial expansion allow to parameterise the $q/p$ of tracks given their entry slope in the dipole magnet and the observed change in slope after traversing the dipole magnet region.}
    \label{fig:polynomailexpansion}
\end{figure}

In order to determine $c_{i=0,1,2,3,4}$ as a function of $t_{x,y}^{\mathrm{Velo}}$, a set of toy tracks are generated. These toy tracks equi-populate the acceptance of $t_{x,y}^{\mathrm{Velo}}$, and have a flat $q/p$ spectrum in each region of $t_{x,y}^{\mathrm{Velo}}$. 
Each parameter $c_{i}$ is fitted with dedicated two dimensional polynomials $c_{i} = \sum_{k,m} c^{km}_{i} \left( t_{x}^{\mathrm{Velo}}\right)^{k}\left(t_{y}^{\mathrm{Velo}}\right)^{m}$. The expansion is done up to a seventh degree ($k+m \leq6$) to ensure a full LHCb acceptance coverage for the parameterisation. The composition of polynomials are done to ensure the $B$ field symmetries are respected selecting only even/odd combinations of $m,k$. The fits to the parameters $c_{i}$ are shown in Figure~\ref{fig:polynomailexpansion}.

\section{Performance}

\begin{figure}[t]
    \centering
    \includegraphics[width=0.45\textwidth]{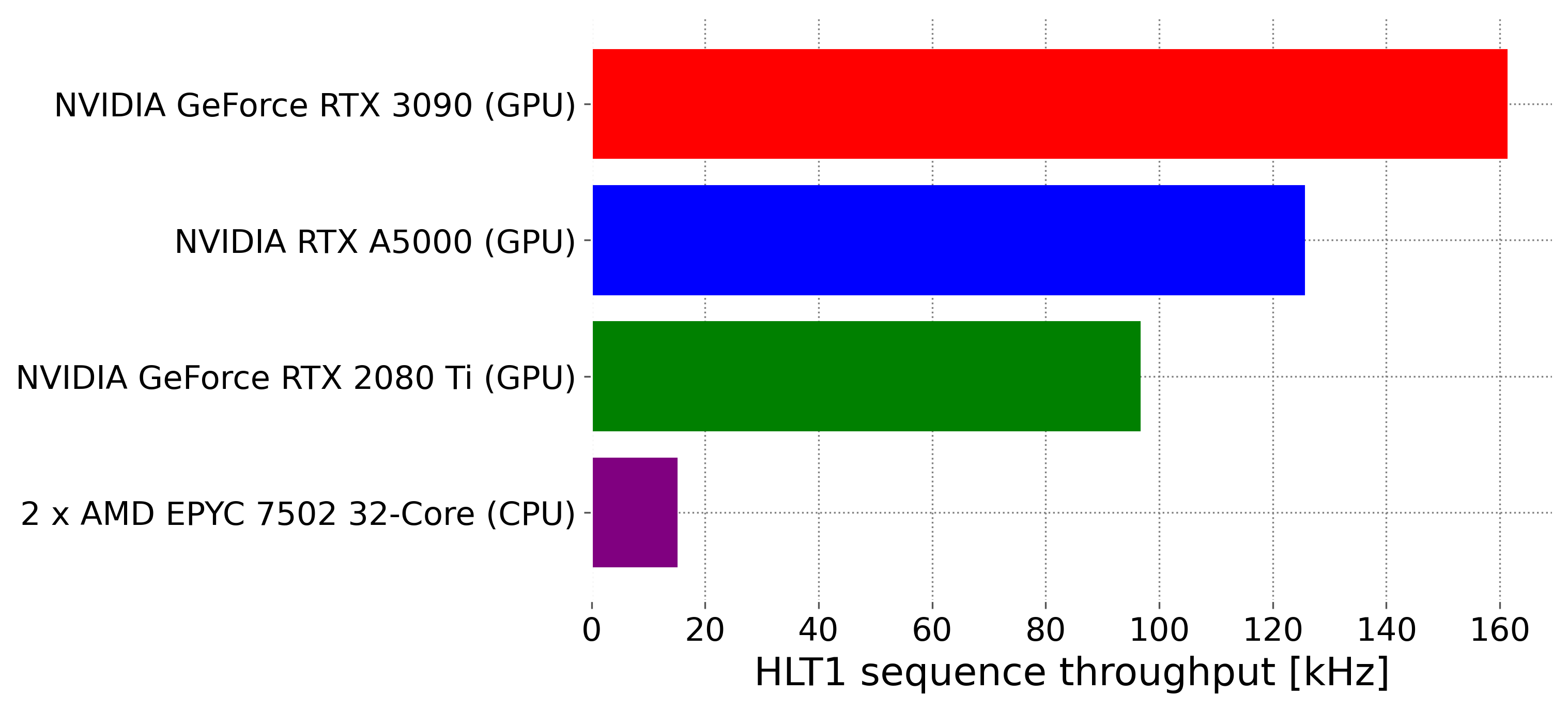}
    \includegraphics[width=0.45\textwidth]{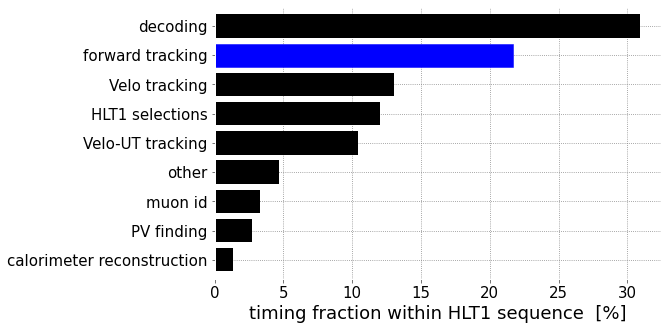}
    \includegraphics[width=0.45\textwidth]{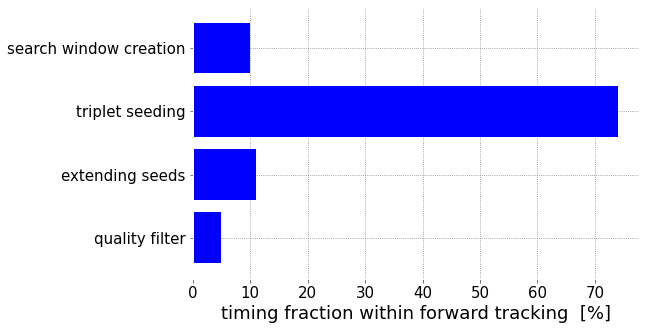}
    \caption{The throughput of the reference HLT1 sequence for three different NVIDIA GPU cards and a dual-socket AMD EPYC 7502 CPU server.
    The breakdown of the HLT1 throughput is shown among the algorithms in the sequence including reconstruction, selection and decoding of the detectors information algorithms. The timing fraction of the substeps of the forward tracking are shown in blue ordered by their execution time. }
    \label{fig:throughput_ref}
\end{figure}

The algorithm is developed and optimised to achieve a tracking efficiency for long tracks with momentum above 5~\gev and \pt above 1~\gev around 90~\% by keeping the overall HLT1 throughput budget per GPU card higher than 105~kHz.
The throughput of the HLT1 sequence is evaluated using a 
sample of simulated minimum bias pp collisions which represent the collisions and detector response in real data. It is shown in 
Figure~\ref{fig:throughput_ref} for both GPU and CPU architectures, as well as the breakdown of this throughput among algorithms. The measured throughput of the HLT1 sequence on a NVIDIA A5000 GPU card is 130~kHz and around six times higher than on AMD EPYC CPU servers. The Looking Forward algorithm employs around 20~\% of the total HLT1 sequence. Within the forward tracking sequence, the largest time fraction is occupied by the triplet seeding step employing a 70~\% fraction of the whole algorithm.

\begin{figure}[t]
    \centering
    \includegraphics[width=0.45\textwidth]{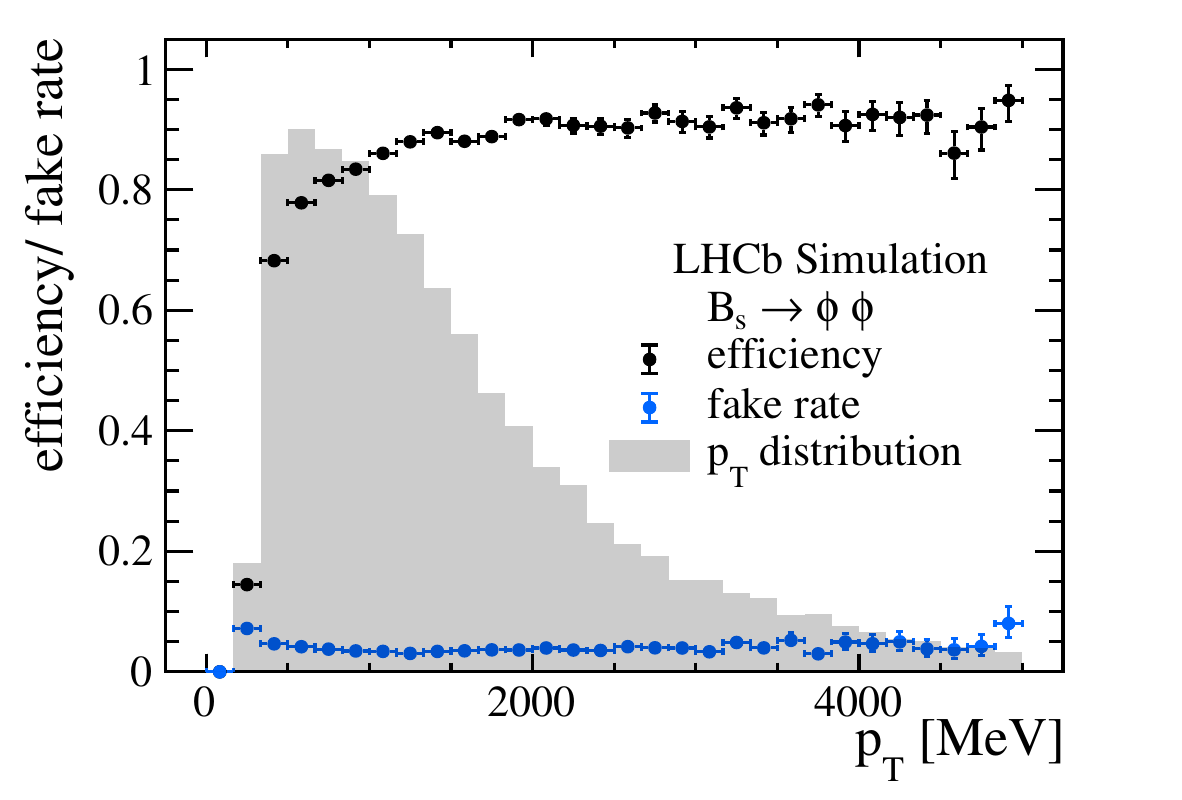}
    \caption{The efficiency and fake rate of the Looking Forward algorithm as tuned for the 
    reference HLT1 sequence, plotted as a function of particle \pt. The different components are described in the figure legend. The distribution of reconstructible charged particles, normalised to unit area, is shown as a shaded histogram to give an idea of the relative physics importance of different kinematic regions.}
    \label{fig:eff_fake_ref_pt}
\end{figure}

\begin{figure}[t]
    \centering
    \includegraphics[width=0.45\textwidth]{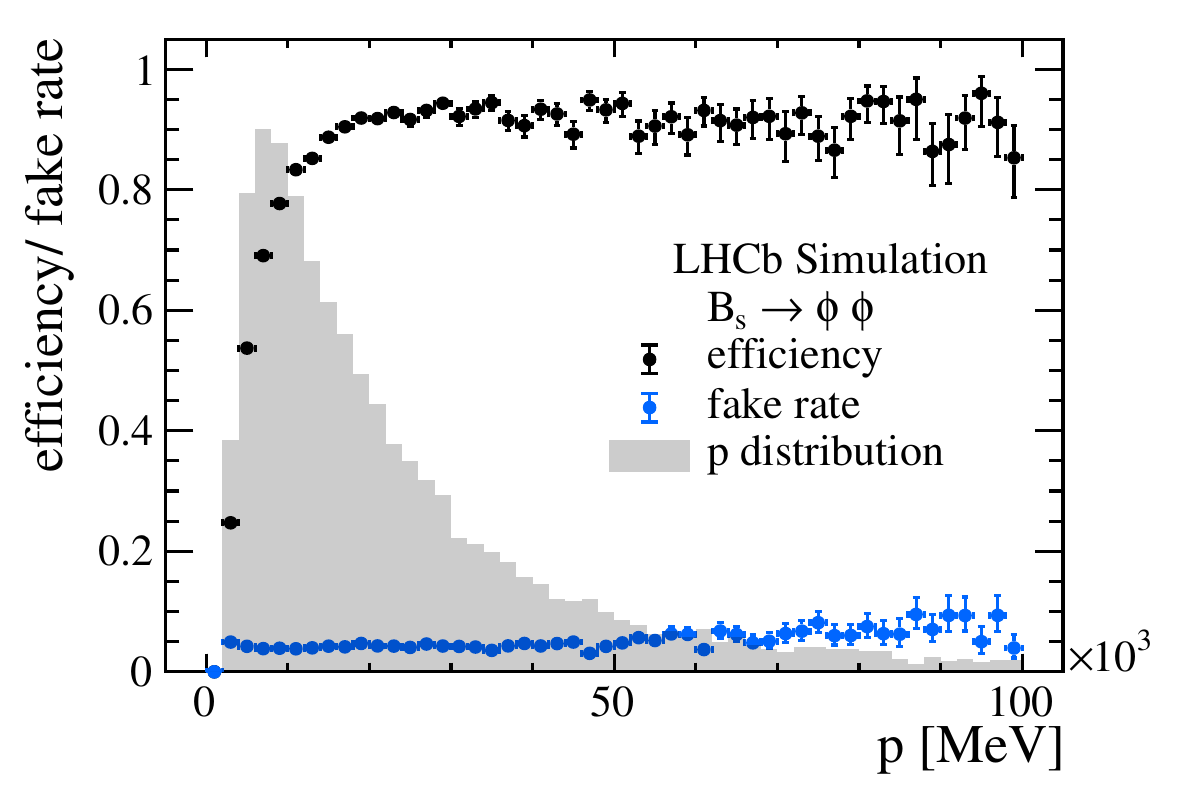}
    \caption{The efficiency and fake rate of the Looking Forward algorithm as tuned for the 
    reference HLT1 sequence, plotted as a function of particle momentum. The different components are described in the figure legend. The distribution of reconstructible charged particles, normalised to unit area, is shown as a shaded histogram to give an idea of the relative physics importance of different kinematic regions.}
    \label{fig:eff_fake_ref_p}
\end{figure}

\begin{figure}[t]
    \centering
    \includegraphics[width=0.45\textwidth]{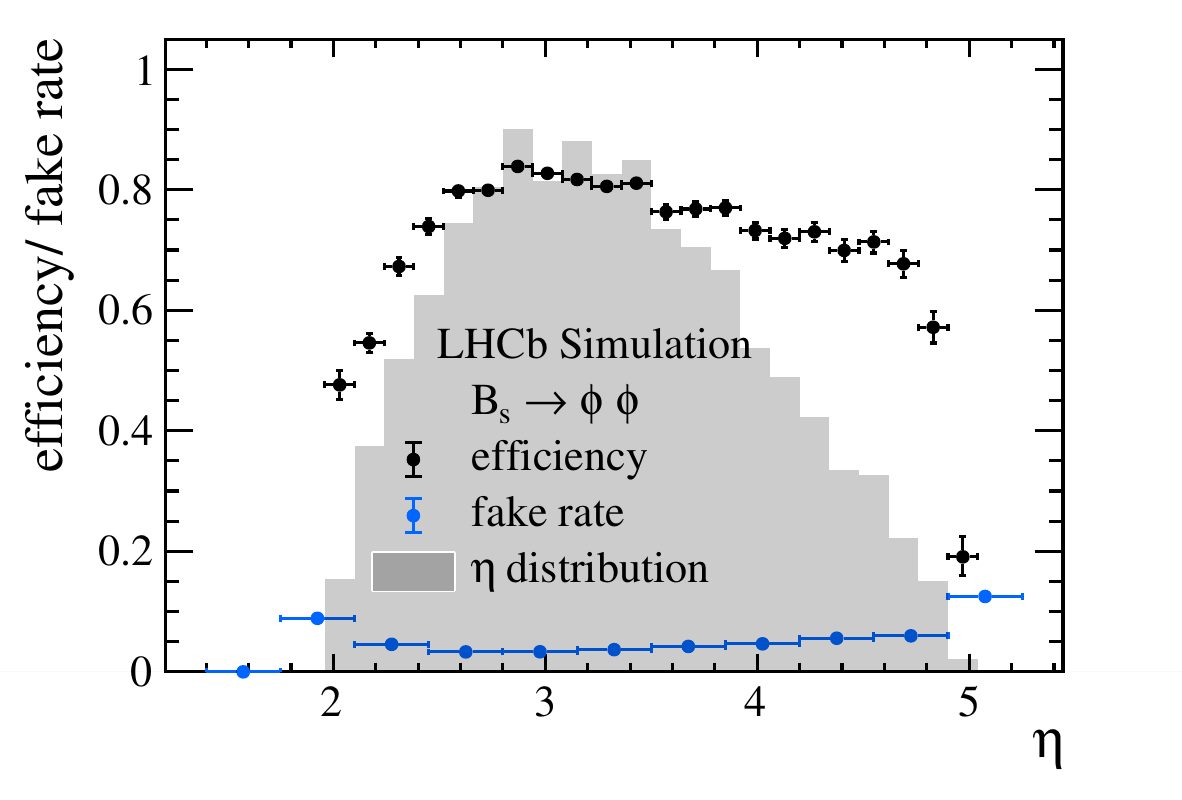}
    \caption{The efficiency and fake rate of the Looking Forward algorithm as tuned for the 
    reference HLT1 sequence, plotted as a function of particle pseudorapidity. The different components are described in the figure legend. The distribution of reconstructible charged particles, normalised to unit area, is shown as a shaded histogram to give an idea of the relative physics importance of different kinematic regions.}
    \label{fig:eff_fake_ref_eta}
\end{figure}

Physics performance is evaluated on a standard sample 
of simulated $B^0_s\to \phi(K^+ K^-)\phi(K^+ K^-)$ decays, which has historically been the decay mode of choice for benchmarking 
the performance of LHCb tracking algorithms. The efficiency for tracks produced in the decays of beauty hadrons, as well as 
the fake rate, is shown in 
Figure~\ref{fig:eff_fake_ref_pt} as a function of particle transverse momentum, Figure~\ref{fig:eff_fake_ref_p} as a function of particle momentum and in Figure~\ref{fig:eff_fake_ref_eta} 
as a function of particle pseudorapidity. The efficiency plateaus is above $90\%$ at 
high transverse momenta or momenta when integrated in the pseudorapidity range $2<\eta<5$. Except at the edges of the algoright 
acceptance, the fake rate is generally flat as a function of both pseudorapidity, transverse momentum and momentum. 

\begin{figure}[t]
    \centering
    \includegraphics[width=0.45\textwidth]{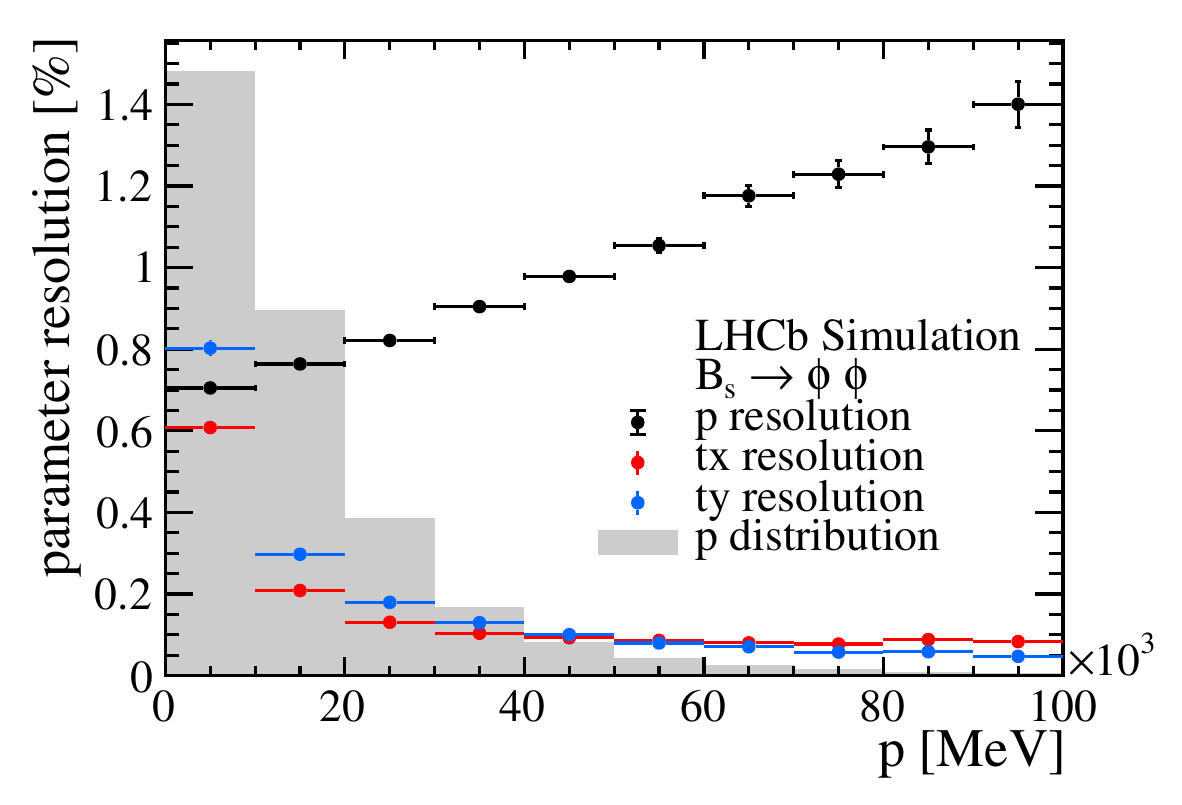}
    \caption{The momentum and track slope at the last SciFi detector layer resolutions of the Looking Forward algorithm as 
    tuned for the reference HLT1 sequence, plotted as a function of particle $p$. The different components are described in the figure legend.}
    \label{fig:res_ref_pt}
\end{figure}

The resolution on track momenta and track slopes is shown in 
Figure~\ref{fig:res_ref_pt} as a function of particle momentum. Resolutions below $1\%$ for both momentum and track slopes are achieved for the great majority of tracks. 

\begin{figure}[t]
    \centering
    \includegraphics[width=0.45\textwidth]{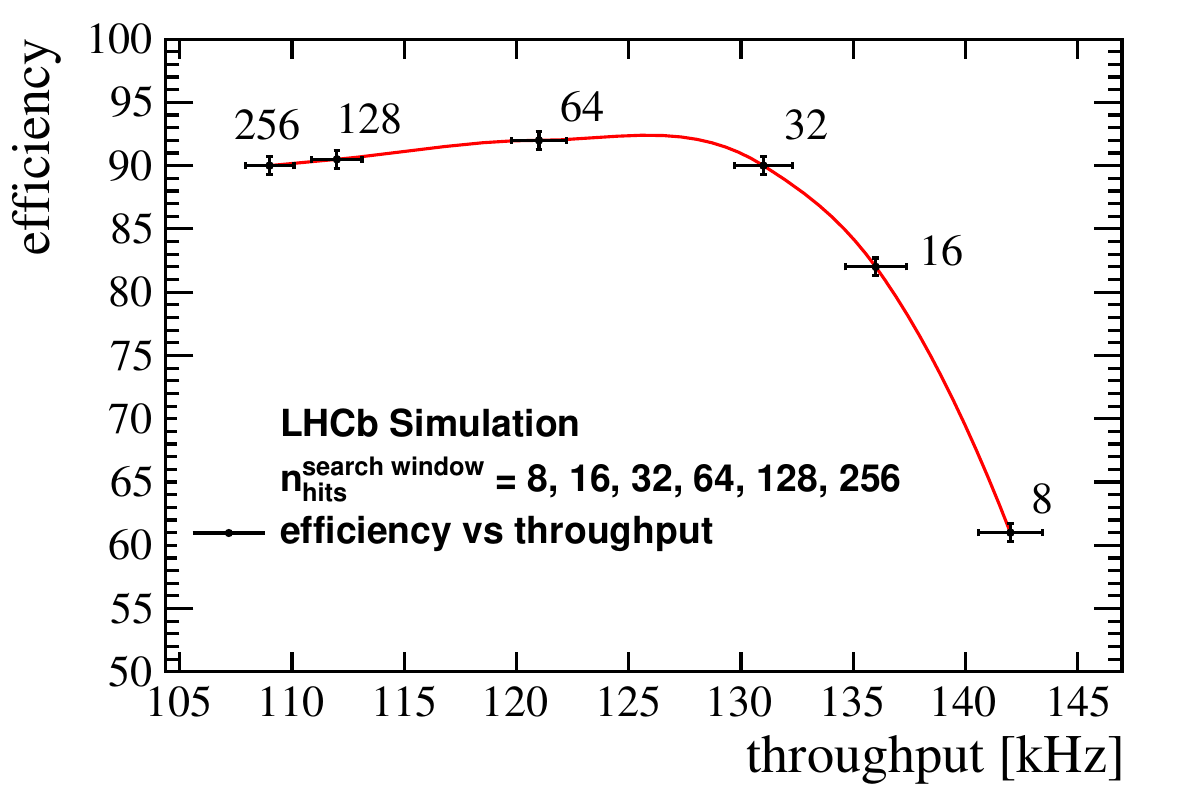}
    \caption{The efficiency and throughput results for different $n_{\mathrm{hits}}^{\mathrm{window}}$ number of hits in the search window values. The $n_{\mathrm{hits}}^{\mathrm{window}}$ variable is proportional to the search window size. The efficiency is evaluated for tracks with momentum above 5~\gev and \pt above 1~\gev and the throughput is measured on a RTX A5000 GPU card using the reference HLT1 sequence.}
    \label{fig:throughput_vs_eff}
\end{figure}

The efficiency and throughput results are shown for different $n_{\mathrm{hits}}^{\mathrm{window}}$ values in Figure~\ref{fig:throughput_vs_eff}. The throughput increases as $n_{\mathrm{hits}}^{\mathrm{window}}$ and the search window size decreases, as less hit combinations are computed, while the total tracking efficiency decreases. If more hit combinations are allowed, enlarging the search window and $n_{\mathrm{hits}}^{\mathrm{window}}$, the efficiency increases however it reaches a plateau above $n_{\mathrm{hits}}^{\mathrm{window}} = 64$ as the algorithm does not achieve the precision required to select the right track candidate among many combinations. The working point of $n_{\mathrm{hits}}^{\mathrm{window}} = 32$ is chosen to the reference as it optimises both efficiency and throughput.

\begin{figure}[t]
    \centering
    \includegraphics[width=0.45\textwidth]{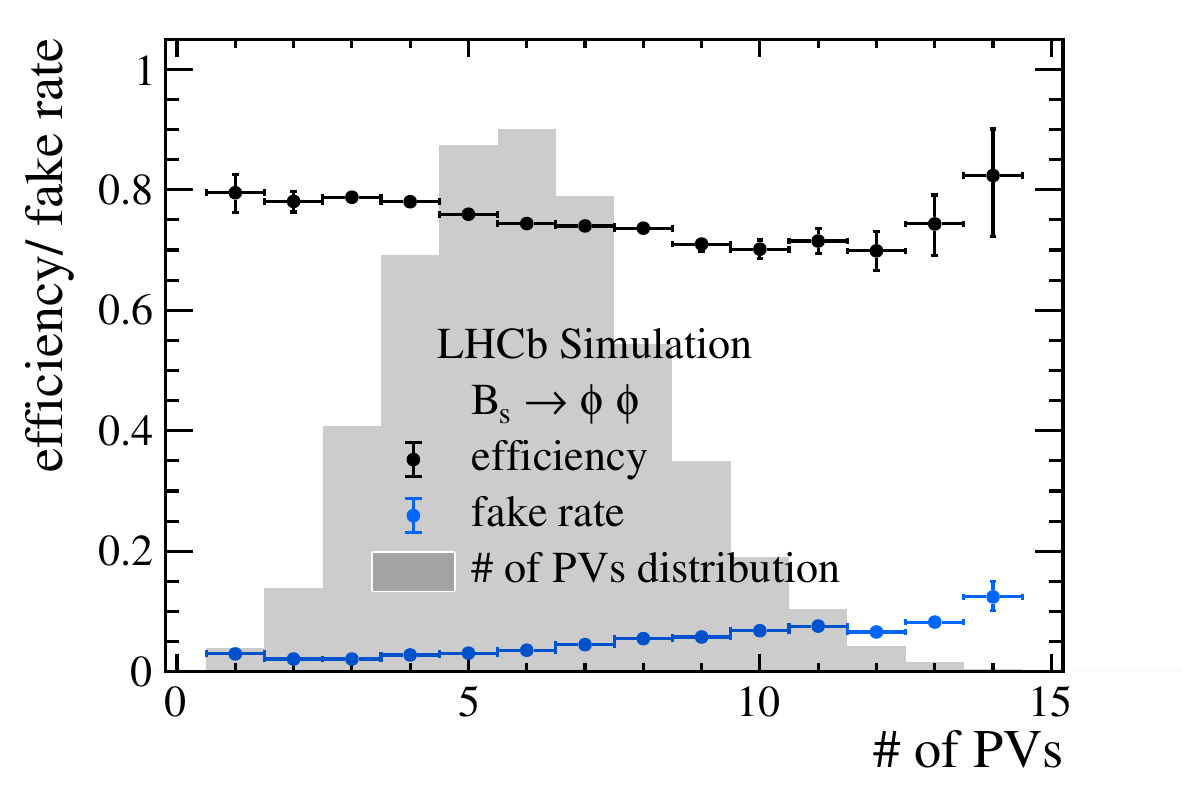}
    \caption{The efficiency and fake rate of the Looking Forward algorithm as tuned for the 
    reference HLT1 sequence, plotted as a function of the number of $pp$ collisions in the event. The different components are 
    described in the figure legend.}
    \label{fig:eff_fake_ref_npv}
\end{figure}

\begin{figure}[t]
    \centering
    \includegraphics[width=0.45\textwidth]{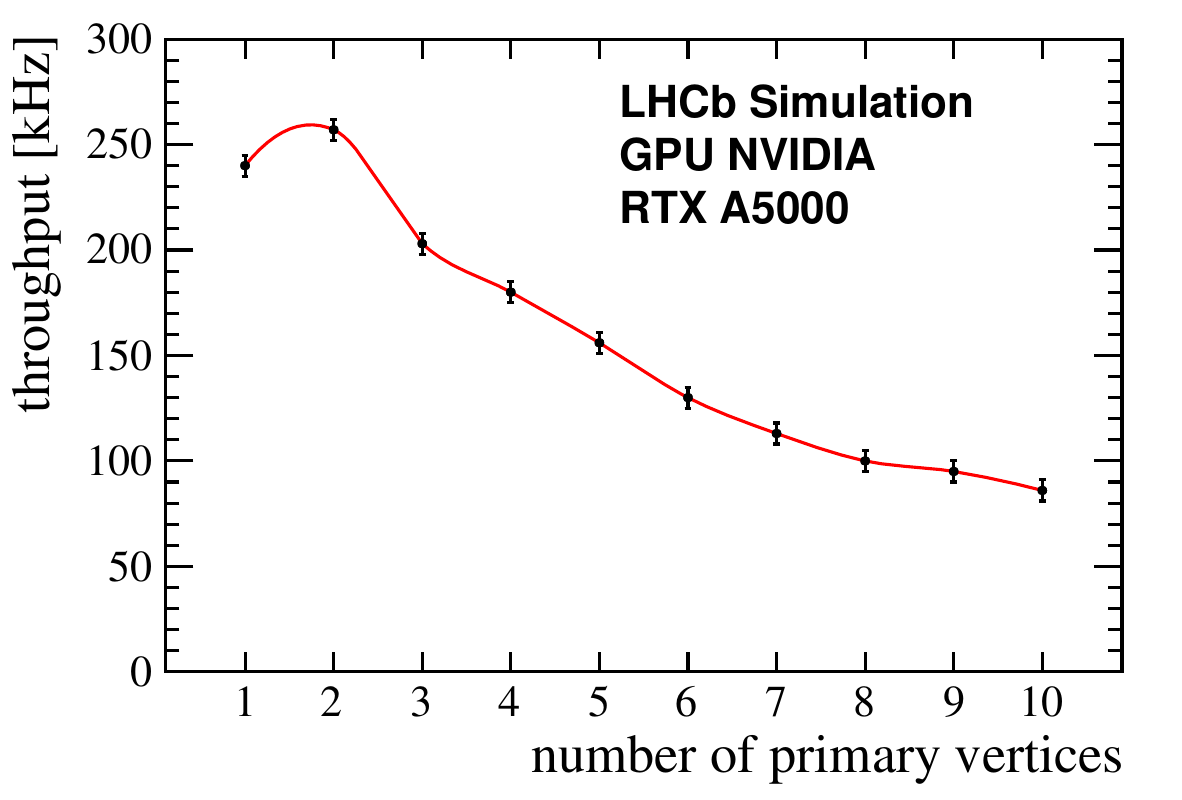}
    \caption{The throughput of the reference HLT1 sequence on the RTX A5000 card as a function of the number of $pp$ collisions in the event.}
    \label{fig:throughput_ref_npv}
\end{figure}

The scalability of this reference configuration is tested by measuring the efficiency and fake rate as a function of the number 
of $pp$ collisions in a simulated event, as shown in Figure~\ref{fig:eff_fake_ref_npv}. While efficiency degrades, and the fake 
rate increases, with an increasing number of collisions, this deterioration is gradual and limited in absolute size. 
In order to benchmark throughput in the same way, dedicated samples of events containing a specific number of $pp$ collisions 
are created from standard LHCb simulation samples. The results are shown in Figure~\ref{fig:throughput_ref_npv}.
The throughput decreases gradually as function of the number of $pp$ collisions, as the number of tracks to be reconstructed in an event increases as a function of it.

We have profiled the forward subsequence of algorithms with the Nsight Compute profiler. The subsequence is dominated by triplet seeding, consisting in 74\% of the subsequence. Triplet seeding has an arithmetic intensity of 29.78, making it compute bound. We observe a low GPU compute throughput of 33\%. The reason behind this inefficiency are stalls in the gangs of threads (warps) scheduled by the CUDA scheduler and can be solved issuing more warps in the kernel call. We are implicitly solving this by running several streams concurrently, which cannot be detected unfortunately by Nsight Compute, as it runs kernels in isolation.

Out of the three parameters that impact occupancy, block size and shared memory are reportedly already optimized for this architecture. However, an area of possible improvement is register utilization, which limits the occupancy of warps to 70\% of what would be theoretically achievable on the A5000 Streaming Multiprocessors. Our kernel requires 72~registers currently, and we are considering reformulations of its critical sections to improve it.

Memory access patterns do not seem to be an issue. We are loading hits in a coalesced manner onto shared memory, and we save tracks upon request by every thread. Triplet seeding only requires fp16 precision for arithmetic, and thus we use fp16 to reduce shared memory utilization in this kernel. As hit data comes in fp32, this results in higher memory pressure than required. If fp16 were reused across other algorithms, it would be sensible to store hit data in fp16 in addition to fp32, however as it stands only triplet seeding can benefit from fp16 and so it is better to pay the arithmetic price once in a non-memory bound kernel. Using fp16 allows us to use \texttt{half-2} vectorized instructions on the GPU, processing two combinations at a time and leading to an efficient formulation that maximizes throughput.

\begin{figure}[t]
    \centering
    \includegraphics[width=0.45\textwidth]{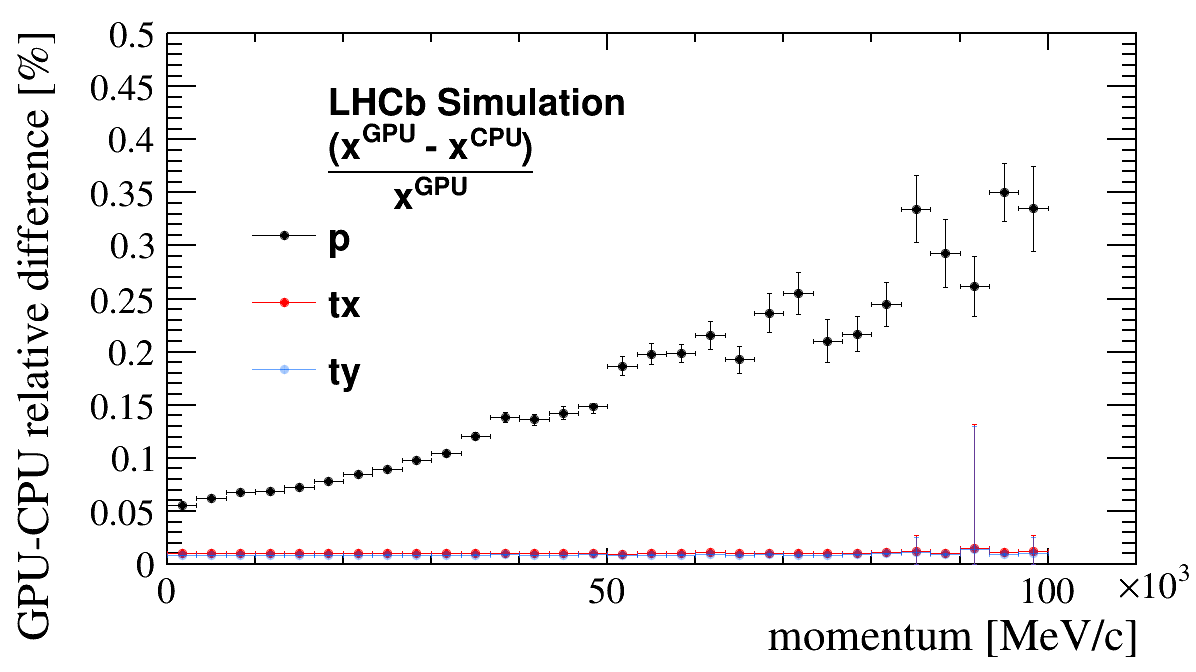}
    \caption{The relative difference between track parameters reconstructed on the A5000 GPU and EPYC 7502 CPU architectures, 
    normalised to the results on the GPU architecture. Around $0.3$\% of tracks are found on only one architecture and are 
    discarded for this comparison. The different parameters which are compared are described in the figure legend.}
    \label{fig:gpucpu_ref}
\end{figure}

The compatibility of physics results on GPU and CPU architectures is investigated by comparing the momentum and slopes of 
reconstructed tracks and is shown in Figure~\ref{fig:gpucpu_ref} for the reference tuning. The momentum relative difference is below 0.3~\% for a wide range of momenta while the track slopes differences around 0.01~\%, matching the per-mille level agreement required, as mentioned in the introduction.  The compatibility is found to be 
insensitive to the specific algorithm tuning and to whether UT information is used or not. 

\begin{figure}[t]
    \centering
    \includegraphics[width=0.45\textwidth]{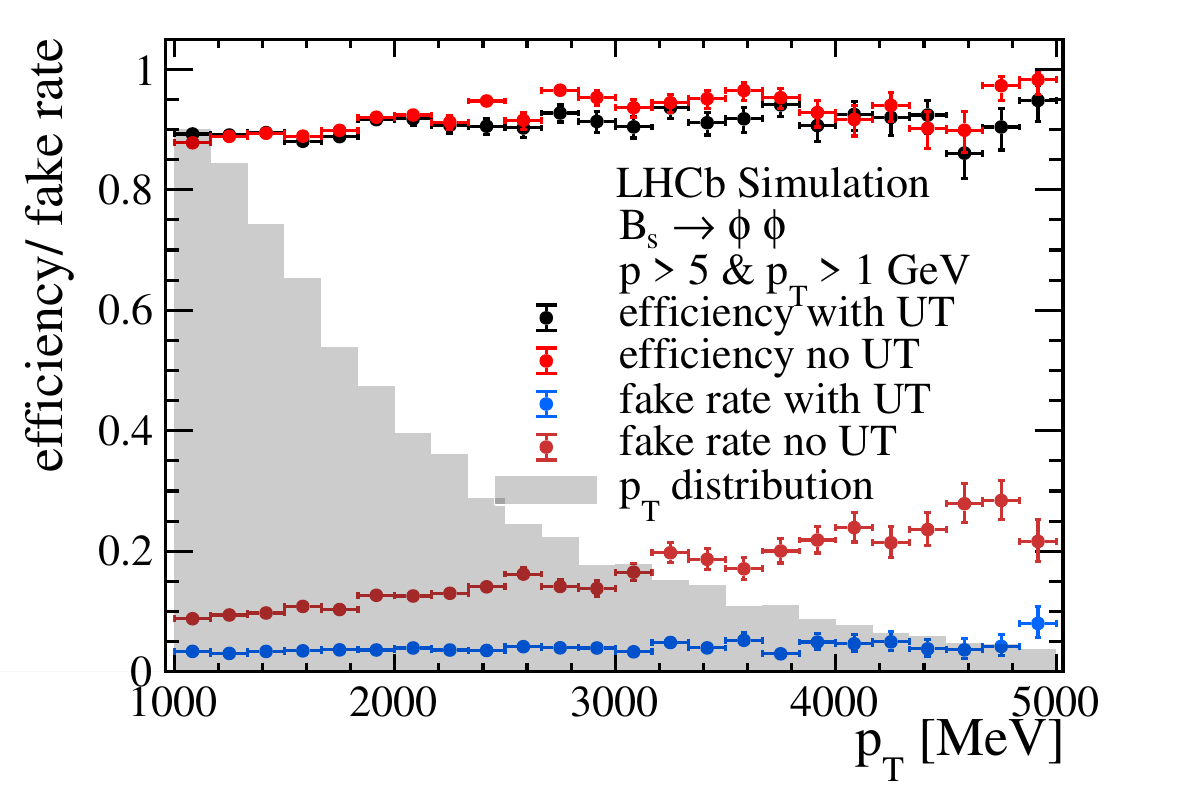}
    \caption{The efficiency and fake rate plotted as a function of particle \pt for the 
    with-UT and no-UT Looking Forward configurations. The different algorithms are described in the figure legend. The distribution of reconstructible charged particles, normalised to unit area, is shown as a shaded histogram to give an idea of the relative physics importance of different kinematic regions.}
    \label{fig:eff_fake_nout_pt}
\end{figure}

\begin{figure}[t]
    \centering
    \includegraphics[width=0.45\textwidth]{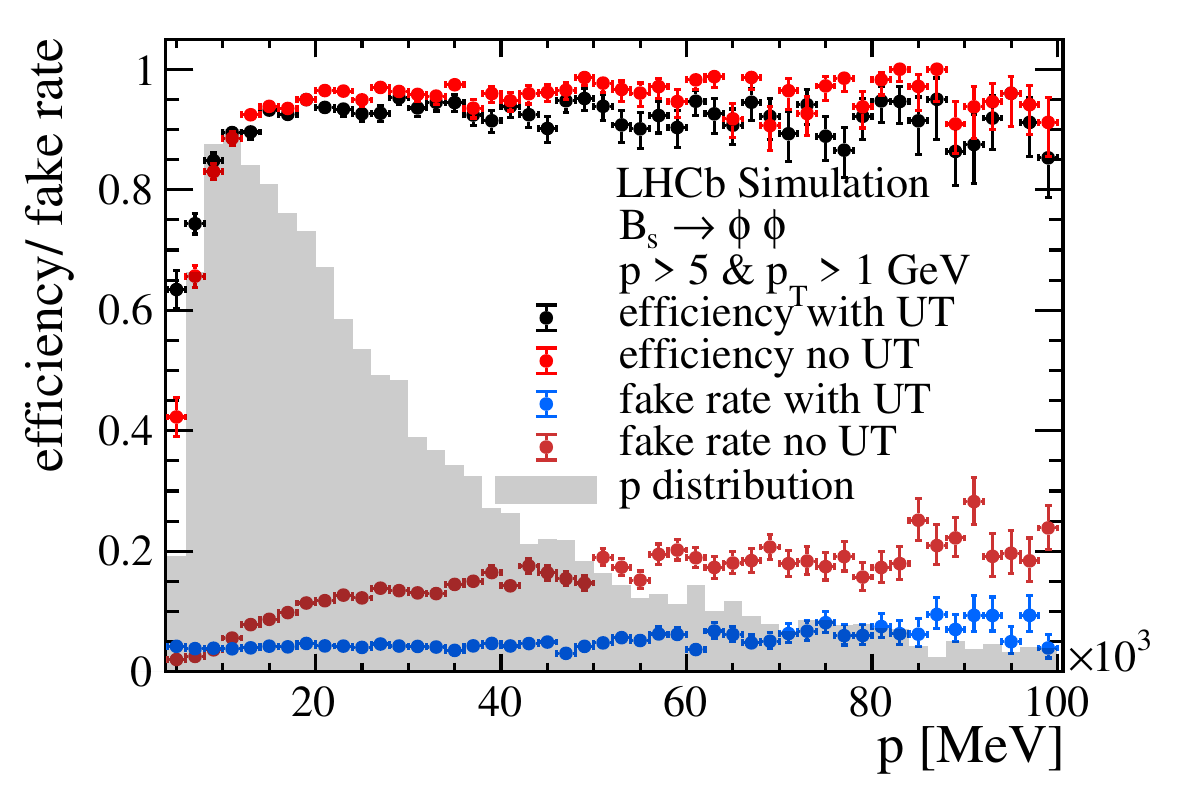}
    \caption{The efficiency and fake rate plotted as a function of particle momentum for the 
    with-UT and no-UT Looking Forward configurations. The different algorithms are described in the figure legend. The distribution of reconstructible charged particles, normalised to unit area, is shown as a shaded histogram to give an idea of the relative physics importance of different kinematic regions.}
    \label{fig:eff_fake_nout_p}
\end{figure}

\begin{figure}[t]
    \centering
    \includegraphics[width=0.45\textwidth]{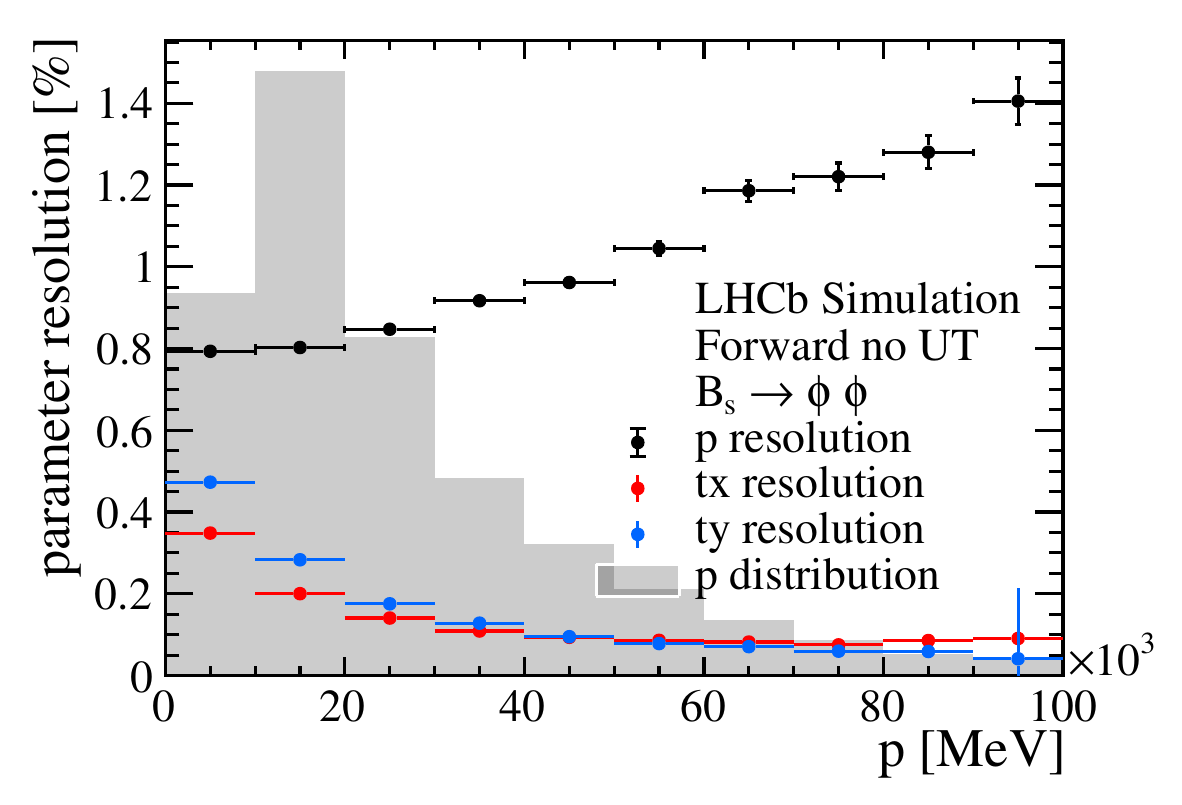}
    \caption{The momentum and track slope at the last SciFi detector layer resolutions of the Looking Forward algorithm as 
    tuned for the no-UT HLT1 sequence, plotted as a function of particle $p$. The different components are described in the figure legend.}
    \label{fig:res_nout_pt}
\end{figure}

The 2022-2023 LHCb data-taking provided an inadvertent test of the Looking Forward algorithm's robustness because the UT detector 
could not be installed in time to participate. It was therefore necessary to reoptimize the algorithm to function without UT 
information, by extrapolating directly from the VELO to the FT as described in the previous section. This logic is analogous to 
that of  the original LHCb forward tracking, used for data-taking from 2009 to 2012, which also extrapolated directly from the 
VELO to the stations downstream of the magnet. It was however thought~\cite{LHCbCollaboration:2014vzo} that such an approach would be impossible for 
HLT1 under Run~3 conditions because of the much greater event rate and number of $pp$ collisions per event. Nevertheless it is
proved possible~\cite{Scarabotto:2823783} to use the algorithm's tunable parameters to achieve an overall throughput of $\sim\! 130$~kHz for 
the HLT1 sequence as a whole, thus validating its fundamental robustness. The Looking Forward ``no-UT'' employs around 40~\% of the HLT1 throughput, twice the amount of the nominal algorithm as more hit combinations are computed due to the lack of the UT information.
The physics performance 
of a ``no-UT'' Looking Forward tuning is compared to that of the reference tuning in Figure~\ref{fig:eff_fake_nout_p} as function of momentum and in Figure~\ref{fig:eff_fake_nout_pt} as a function of \pt. The 
fake rate roughly triples for the same efficiency, which is expected since the UT hits are essential in discriminating 
between correct and false matches of VELO and FT track segments. Nevertheless the fake rate remains below $15\%$ for 
most tracks. The momentum and track slopes resolutions are shown in Figure~\ref{fig:res_nout_pt} which can be compared to Figure~\ref{fig:res_ref_pt} 
and shows only a modest degradation.

\begin{figure}[t]
    \centering
    \includegraphics[width=0.45\textwidth]{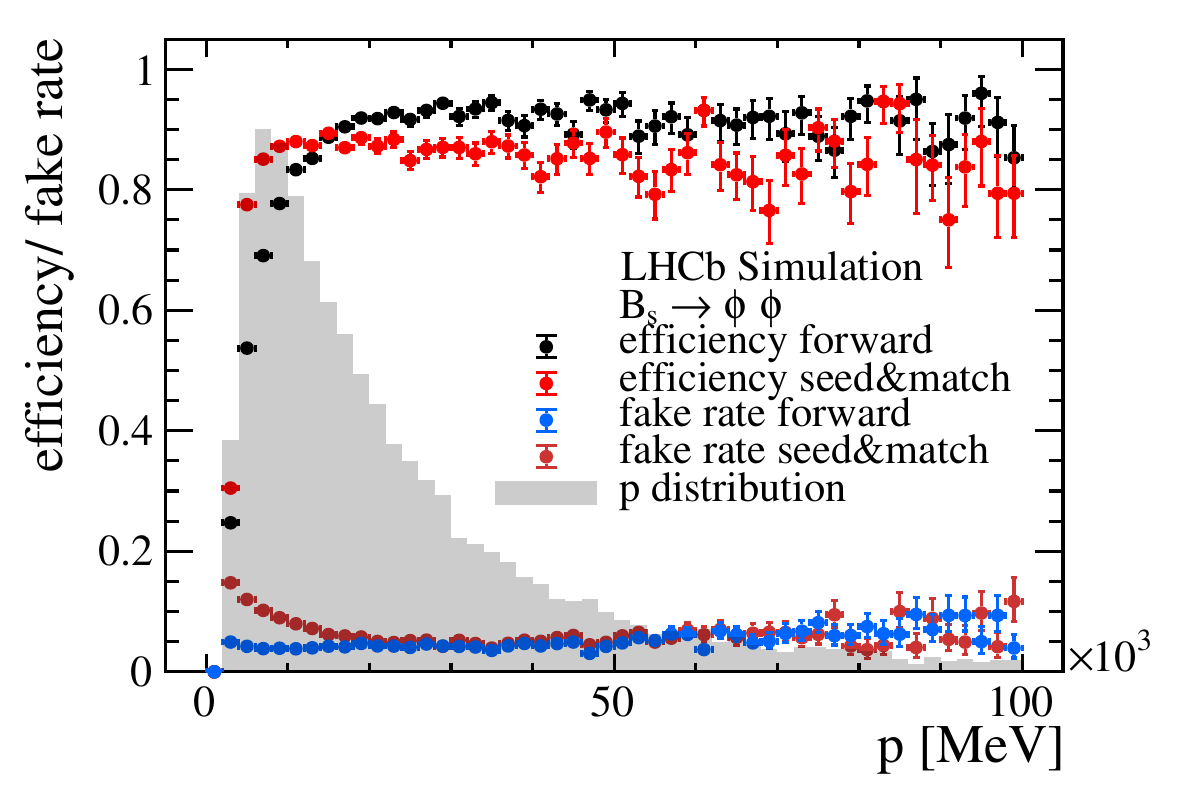}
    \caption{The efficiency and fake rate plotted as a function of particle momentum for the forward and seeding $\&$ matching configurations. The different algorithms are described in the figure legend. The distribution of reconstructible charged particles, normalised to unit area, is shown as a shaded histogram to give an idea of the relative physics importance of different kinematic regions.}
    \label{fig:eff_fake_forward_vs_match}
\end{figure}

To exploit the second GPU card installed by LHCb in 2023, tradeoffs between computational cost and physics performance are studied in HLT1 by
implementing a best long track reconstruction similarly on how tracks are reconstructed in HLT2~\cite{Gunther:2865000}. This algorithm first reconstruct tracks with the Looking Forward method, then flags all the unused hits in the detector and finally reconstruct with SciFi seeding and matching GPU-optimised strategy~\cite{Calefice:2856339}. As shown in Figure~\ref{fig:eff_fake_forward_vs_match}, the seeding and matching method performs better at lower momenta while the forward one at higher momenta. The best long track reconstruction tries to exploit the advantages of both approaches.

\begin{figure}[t]
    \centering
    \includegraphics[width=0.45\textwidth]{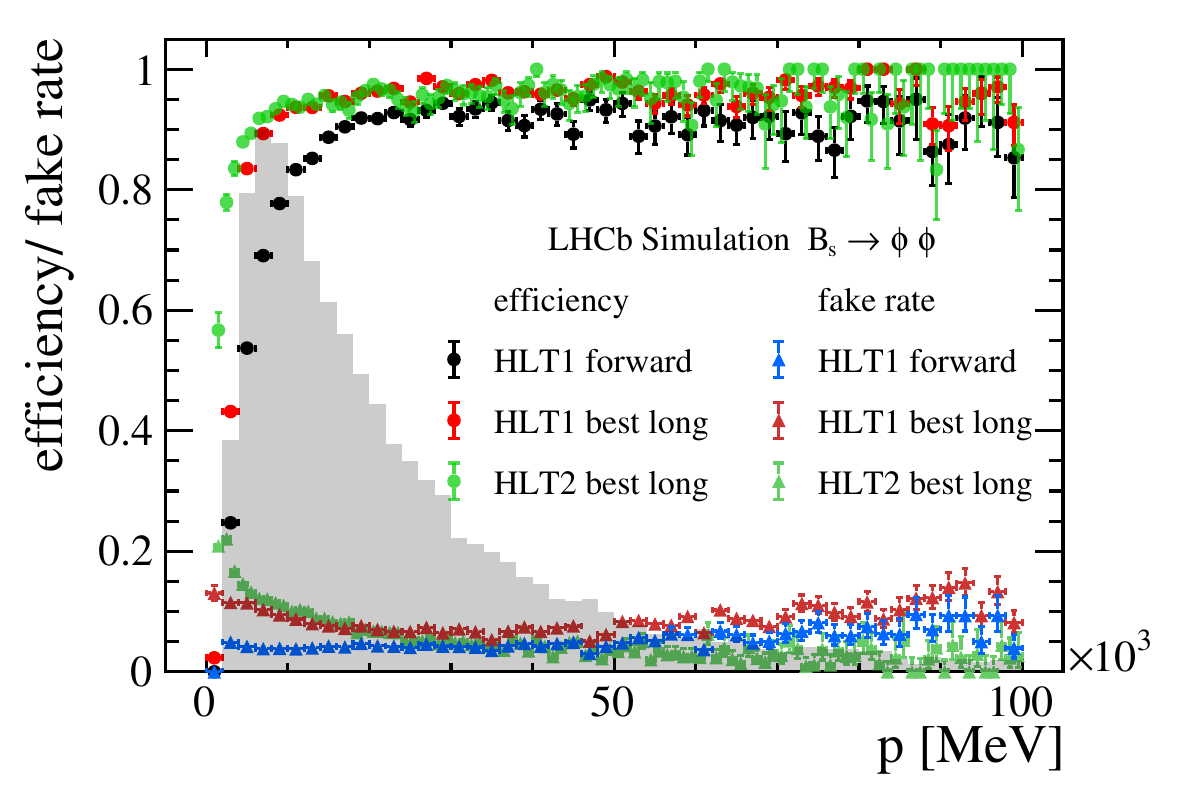}
    \caption{The efficiency and fake rate plotted as a function of momentum of particles for the 
    forward and best long track reconstruction in HLT1 and HLT2. The different algorithms are described in the figure legend. The distribution of reconstructible charged particles, normalised to unit area, is shown as a shaded histogram to give an idea of the relative physics importance of different kinematic regions.}
    \label{fig:eff_fake_loose_pt}
\end{figure}

The overall HLT1 sequence throughput is reduced by around 30~\% reaching 90~kHz, with all the additional resource usage allocated to the long track reconstruction. This is an extreme scenario and not necessarily 
representative of how the collaboration will use its resources, but rather provides an upper bound on these tradeoffs. 
Figure~\ref{fig:eff_fake_loose_pt} shows the efficiency and fake rate as a function of particle momentum, comparing the 
Looking Forward algorithm with the best long track reconstruction configurations implemented in HLT1 and HLT2. The best long track sequence improves the tracking efficiency of the HLT1 reconstruction at low momentum. The long tracking efficiency of HLT2 remains higher as lower momentum requirements are applied when performing the reconstruction and a full Kalman filter is exploited to improve the track resolution.

\section{Prospects and conclusions}

We have presented Looking Forward, a new algorithm exploiting the track following approach and optimised on parallel GPU architectures.
We developed the algorithm in order to maximise the throughput while achieving the best physics performance. 

The method extrapolates tracks reconstructed before the LHCb dipole magnet to SciFi tracker after the magnet. The algorithm parellalises the search for hits in SciFi detector over the input tracks. The hits found in this way are combined in parallel to form candidate tracklets, which are then combined to the input tracks. The method achieves a 90~\% tracking efficiency across a large spectrum of momenta and a momentum resolution below 1~\%. The measured throughput on a A5000 card is $\sim\!\!\! 130$~kHz.

The algorithm features tunable parameters which can be adapted on the physics requirements. The absence of the UT subdetector during the 2022 data-taking provided a stress test to the algorithm which was adapted to handle the missing information. The method in such configuration maintains the physics and computational performances in a sub-range of tracks with momentum greater than 5 \gevc and transverse momentum greater than 1 \gevc.

The algorithm was included and commissioned during the 2022 LHCb data-taking, and planned to be used in production in the coming years. We will continue exploring techniques to obtain better performances in the current and upcoming hardware generations.

\section*{Acknowledgements}
We thank LHCb's Real-Time Analysis project for its support, for many useful discussions, and for reviewing an early draft of this manuscript. We also thank the LHCb computing and simulation teams for producing the simulated LHCb samples used to benchmark the performance of the algorithm presented in this paper. The development and maintenance of LHCb's nightly testing and benchmarking infrastructure which our work relied on is a collaborative effort and we are grateful to all LHCb colleagues who contribute to it. VVG and AS are supported by the European Research Council under Grant Agreement number 724777 ``RECEPT''. DvB acknowledges support of the European Research Council Starting grant ALPACA 101040710. 
\EOD

\printbibliography[heading=bibintoc]

@article{Dorenbosch:1985cx,
    author = "Dorenbosch, Jheroen",
    title = "{Trigger in UA2 and in UA1}",
    reportNumber = "NIKHEF-H-85-14",
    journal = "eConf",
    volume = "C851111",
    pages = "134--151",
    year = "1985"
}

@article{Aiola:2020ydy,
    author = "Aiola, Salvatore and others",
    title = "{Hybrid seeding: A standalone track reconstruction algorithm for scintillating fibre tracker at LHCb}",
    eprint = "2007.02591",
    archivePrefix = "arXiv",
    primaryClass = "physics.ins-det",
    doi = "10.1016/j.cpc.2020.107713",
    journal = "Comput. Phys. Commun.",
    volume = "260",
    pages = "107713",
    year = "2021"
}

@article{Benayoun:2002hqa,
    author = "Benayoun, M. and Callot, O.",
    title = "{The forward tracking, an optical model method}",
    reportNumber = "LHCb-2002-008, CERN-LHCb-2002-008",
    month = "2",
    year = "2002"
}

@article{LHCb:2023hlw,
    author = "Aaij, Roel and others",
    collaboration = "LHCb",
    title = "{The LHCb upgrade I}",
    eprint = "2305.10515",
    archivePrefix = "arXiv",
    primaryClass = "hep-ex",
    reportNumber = "LHCb-DP-2022-002",
    month = "5",
    year = "2023"
}

@techreport{Bowen:1635665,
      author        = "Bowen, E and Storaci, B",
      title         = "{VeloUT tracking for the LHCb Upgrade}",
      institution   = "CERN",
      reportNumber  = "LHCb-PUB-2013-023, CERN-LHCb-PUB-2013-023,
                       LHCb-INT-2013-056",
      address       = "Geneva",
      year          = "2014",
      url           = "https://cds.cern.ch/record/1635665",
}

@thesis{Quagliani:2296404,
      author        = "Quagliani, Renato",
      title         = "{Study of double charm B decays with the LHCb experiment
                       at CERN and track reconstruction for the LHCb upgrade}",
      year          = "2017",
      url           = "https://cds.cern.ch/record/2296404",
}

@thesis{Hasse:2706588,
      author        = "Hasse, Christoph",
      title         = "{Alternative approaches in the event reconstruction of
                       LHCb}",
      year          = "2019",
      url           = "https://cds.cern.ch/record/2706588",
}

@article{Kutta,
      author        = "Kutta, W",
      title         = "{Beitrag zur näherungsweisen Integration totaler Differentialgleichungen}",
      year          = "1901",
	  journal={Z. Math.Phys.}, 
	volume={46},
	pages={435-453},
}

@thesis{Calefice:2856339,
      author        = "Calefice, Lukas",
      title         = "{Standalone track reconstruction on GPUs in the first
                       stage of the upgraded LHCb trigger system and
                       Preparations for measurements with strange hadrons in Run
                       3}",
      year          = "2022",
      url           = "https://cds.cern.ch/record/2856339",
}

@article{FernandezDeclara:2019ycx,
    author = "Fernandez Declara, Placido and Campora Perez, Daniel Hugo and Vom Bruch, Dorothea and Neufeld, Niko and Garcia-Blas, Javier and Daniel Garcia, J.",
    title = "{A Parallel-Computing Algorithm for High-Energy Physics Particle Tracking and Decoding Using GPU Architectures}",
    eprint = "2002.11529",
    archivePrefix = "arXiv",
    primaryClass = "physics.ins-det",
    doi = "10.1109/ACCESS.2019.2927261",
    journal = "IEEE Access",
    volume = "7",
    pages = "91612--91626",
    year = "2019"
}

@article{CamporaPerez:2021jhc,
    author = "C\'ampora P\'erez, Daniel Hugo and Neufeld, Niko and Riscos N\'u\~nez, Agust\'\i{}n",
    title = "{Search by triplet: An efficient local track reconstruction algorithm for parallel architectures}",
    eprint = "2207.03936",
    archivePrefix = "arXiv",
    primaryClass = "hep-ex",
    doi = "10.1016/j.jocs.2021.101422",
    journal = "J. Comput. Sci.",
    volume = "54",
    pages = "101422",
    year = "2021"
}

@techreport{Li:2752971,
      author        = "Li, Peilian and Rodrigues, Eduardo and Stahl, Sascha",
      title         = "{Tracking Definitions and Conventions for Run 3 and
                       Beyond}",
      institution   = "CERN",
      reportNumber  = "LHCb-PUB-2021-005, CERN-LHCb-PUB-2021-005",
      address       = "Geneva",
      year          = "2021",
      url           = "https://cds.cern.ch/record/2752971",
}

@article{LHCbCollaboration:2014vzo,
    title = "{LHCb Trigger and Online Upgrade Technical Design Report}",
    reportNumber = "CERN-LHCC-2014-016, LHCB-TDR-016",
    month = "5",
    year = "2014"
}

@article{Scarabotto:2823783,
      author        = "Scarabotto, Alessandro",
      title         = "{Tracking on GPU at LHCb’s fully software trigger}",
      year          = "2022",
      url           = "https://cds.cern.ch/record/2823783",
}

@techreport{LHCbCollaboration:2717938,
      author        = "LHCb Collaboration, CERN (Meyrin)",
      title         = "{LHCb Upgrade GPU High Level Trigger Technical Design
                       Report}",
      institution   = "CERN",
      reportNumber  = "CERN-LHCC-2020-006, LHCB-TDR-021",
      address       = "Geneva",
      year          = "2020",
      url           = "https://cds.cern.ch/record/2717938",
      doi           = "10.17181/CERN.QDVA.5PIR",
}

@article{Aaij:2019zbu,
    author = "Aaij, Roel and others",
    title = "{Allen: A high level trigger on GPUs for LHCb}",
    eprint = "1912.09161",
    archivePrefix = "arXiv",
    primaryClass = "physics.ins-det",
    doi = "10.1007/s41781-020-00039-7",
    journal = "Comput. Softw. Big Sci.",
    volume = "4",
    number = "1",
    pages = "7",
    year = "2020"
}

@article{LHCb:2014nio,
    author = "Aaij, Roel and others",
    collaboration = "LHCb",
    title = "{Measurement of the track reconstruction efficiency at LHCb}",
    eprint = "1408.1251",
    archivePrefix = "arXiv",
    primaryClass = "hep-ex",
    reportNumber = "CERN-LHCB-DP-2013-002",
    doi = "10.1088/1748-0221/10/02/P02007",
    journal = "JINST",
    volume = "10",
    number = "02",
    pages = "P02007",
    year = "2015"
}

@article{LHCb:2019gvd,
    author = "Aaij, Roel and others",
    collaboration = "LHCb",
    title = "{Measurement of the electron reconstruction efficiency at LHCb}",
    eprint = "1909.02957",
    archivePrefix = "arXiv",
    primaryClass = "hep-ex",
    reportNumber = "LHCb-DP-2019-003, CERN-EP-2019-181",
    doi = "10.1088/1748-0221/14/11/P11023",
    journal = "JINST",
    volume = "14",
    pages = "P11023",
    year = "2019"
}

@ARTICLE{1018421,
  author={Andre, J. and Charra, P. and Flegel, W. and Giudici, P.A. and Jamet, O. and Lancon, P. and Losasso, M. and Rohner, F. and Rosset, C.},
  journal={IEEE Transactions on Applied Superconductivity}, 
  title={Status of the LHCb magnet system}, 
  year={2002},
  volume={12},
  number={1},
  pages={366-371},
  doi={10.1109/TASC.2002.1018421}}

@ARTICLE{1324843,
  author={Andre, J. and Flegel, W. and Giudici, P.A. and Jamet, O. and Losasso, M.},
  journal={IEEE Transactions on Applied Superconductivity}, 
  title={Status of the LHCb dipole magnet}, 
  year={2004},
  volume={14},
  number={2},
  pages={509-513},
  doi={10.1109/TASC.2004.829705}}

@article{Colombo:2019bel,
    author = "Colombo, Tommaso and Durante, Paolo and Galli, Domenico and Marconi, Umberto and Neufeld, Niko and Pisani, Flavio and Schwemmer, Rainer and Valat, S\'ebastien and Voneki, Balazs",
    title = "{The LHCb DAQ Upgrade for LHC Run3}",
    doi = "10.1109/TNS.2019.2920393",
    journal = "IEEE Trans. Nucl. Sci.",
    volume = "66",
    number = "7",
    pages = "982--985",
    year = "2019"
}

@article{ATLAS:2023dns,
    author = "Aad, Georges and others",
    collaboration = "ATLAS",
    title = "{The ATLAS Experiment at the CERN Large Hadron Collider: A Description of the Detector Configuration for Run 3}",
    eprint = "2305.16623",
    archivePrefix = "arXiv",
    primaryClass = "physics.ins-det",
    reportNumber = "CERN-EP-2022-259",
    month = "5",
    year = "2023"
}

@article{CMS-DP-2023-028,
      collaboration = "CMS",
      title         = "{Performance of Track Reconstruction at the CMS High-Level
                       Trigger in 2022 data}",
      year          = "2023",
      url           = "https://cds.cern.ch/record/2860207",
}

@article{ATLAS:2021vfj,
    collaboration = "ATLAS",
    title = "{Software Performance of the ATLAS Track Reconstruction for LHC Run 3}",
    institution   = "CERN",
      reportNumber  = "ATL-PHYS-PUB-2021-012",
      address       = "Geneva",
      year          = "2021",
      url           = "https://cds.cern.ch/record/2766886",
     
}

@article{Antonioli:2013ppp,
    editor = "Antonioli, P. and Kluge, A. and Riegler, W.",
    collaboration = "ALICE",
    title = "{Upgrade of the ALICE Readout $\&$ Trigger System}",
    reportNumber = "CERN-LHCC-2013-019, ALICE-TDR-015",
    year = "2013"
}

@techreport{LHCbCollaboration:2776420,
      author        = "LHCb Collaboration, CERN (Meyrin)",
      title         = "{Framework TDR for the LHCb Upgrade II -
                       Opportunities in flavour physics, and beyond, in the HL-LHC
                       era}",
      institution   = "CERN",
      reportNumber  = "CERN-LHCC-2021-012, LHCB-TDR-023",
      address       = "Geneva",
      year          = "2021",
      url           = "https://cds.cern.ch/record/2776420",
}

@techreport{Collaboration:2759072,
      author        = "Collaboration, CMS",
      title         = "{The Phase-2 Upgrade of the CMS Data Acquisition and High
                       Level Trigger}",
      institution   = "CERN",
      reportNumber  = "CERN-LHCC-2021-007, CMS-TDR-022",
      address       = "Geneva",
      year          = "2021",
      url           = "https://cds.cern.ch/record/2759072",
      $note          = "This is the final version of the document, approved by the
                       LHCC",
}

@techreport{ATLAS:2802799,
      author        = "ATLAS, Collaboration",
      title         = "{Technical Design Report for the Phase-II Upgrade of the
                       ATLAS Trigger and Data Acquisition System - Event Filter
                       Tracking Amendment}",
      institution   = "CERN",
      reportNumber  = "CERN-LHCC-2022-004, ATLAS-TDR-029-ADD-1",
      address       = "Geneva",
      year          = "2022",
      url           = "https://cds.cern.ch/record/2802799",
      doi           = "10.17181/CERN.ZK85.5TDL",
}

@article{LHCb:2018zdd,
    author = "Aaij, Roel and others",
    collaboration = "LHCb",
    title = "{Design and performance of the LHCb trigger and full real-time reconstruction in Run 2 of the LHC}",
    eprint = "1812.10790",
    archivePrefix = "arXiv",
    primaryClass = "hep-ex",
    reportNumber = "CERN-LHCb-DP-2019-001",
    doi = "10.1088/1748-0221/14/04/P04013",
    journal = "JINST",
    volume = "14",
    number = "04",
    pages = "P04013",
    year = "2019"
}

@article{ATLAS:2020esi,
    author = "Aad, Georges and others",
    collaboration = "ATLAS",
    title = "{Operation of the ATLAS trigger system in Run 2}",
    eprint = "2007.12539",
    archivePrefix = "arXiv",
    primaryClass = "physics.ins-det",
    reportNumber = "CERN-EP-2020-109",
    doi = "10.1088/1748-0221/15/10/P10004",
    journal = "JINST",
    volume = "15",
    number = "10",
    pages = "P10004",
    year = "2020"
}

@article{CMS:2016ngn,
    author = "Khachatryan, Vardan and others",
    collaboration = "CMS",
    title = "{The CMS trigger system}",
    eprint = "1609.02366",
    archivePrefix = "arXiv",
    primaryClass = "physics.ins-det",
    reportNumber = "CMS-TRG-12-001, CERN-EP-2016-160",
    doi = "10.1088/1748-0221/12/01/P01020",
    journal = "JINST",
    volume = "12",
    number = "01",
    pages = "P01020",
    year = "2017"
}

@thesis{Gunther:2865000,
      author        = "Gunther, Andre",
      title         = "{Track Reconstruction Development and Commissioning for
                       LHCb's  Run 3 Real-time Analysis Trigger}",
      year          = "2023",
      url           = "https://cds.cern.ch/record/2865000",
}

@article{Callot:2007mba,
    author = "Callot, O. and Hansmann-Menzemer, S.",
    title = "{The Forward Tracking}: {Algorithm and Performance Studies}",
    reportNumber = "LHCb-2007-015, CERN-LHCb-2007-015",
    month = "5",
    year = "2007"
}

@article{CMS:2023gfb,
    author = "Hayrapetyan, Aram and others",
    collaboration = "CMS",
    title = "{Development of the CMS detector for the CERN LHC Run 3}",
    eprint = "2309.05466",
    archivePrefix = "arXiv",
    primaryClass = "physics.ins-det",
    reportNumber = "CMS-PRF-21-001, CERN-EP-2023-136",
    month = "9",
    year = "2023"
}

\end{document}